\documentclass[letterpape,11pt]{article}
\pdfoutput=1 

\usepackage{jheppub}

\usepackage{amsmath}
\usepackage{slashed}
\usepackage{amssymb}
\usepackage{mathtools}
\usepackage{graphicx}
\usepackage{subfigure}
\usepackage{setspace}
\usepackage{sidecap}
\usepackage{color}
\usepackage{hyperref}
\usepackage{tabu} 
\usepackage{hyperref}
\usepackage[export]{adjustbox}
\usepackage{tensor}

\newcommand{\lb}{\Big{\lbrack}}
\newcommand{\rb}{\Big{\rbrack}}
\newcommand{\lp}{\Big{(}}
\newcommand{\rp}{\Big{)}}
\newcommand{\lbc}{\Big{\lbrace}}
\newcommand{\rbc}{\Big{\rbrace}}
\newcommand{\nn}{\nonumber}
\newcommand{\Bvert}{\Big{\vert}}
\newcommand{\Rangle}{\Big{\rangle}}
\newcommand{\Langle}{\Big{\langle}}
\newcommand{\bmat}[1]{{\boldsymbol{\mathrm{#1}}}}

\newcommand{\Q}[4]{ {}^{#1} #2 ^{[#4]}_{#3} }
\newcommand{\ve}{\lambda}

\title{An Effective Theory of Quarkonia in QCD Matter }
\date{\today}
\author[a]{Yiannis Makris}
\author[a]{and Ivan Vitev}
\affiliation[a]{Theoretical Division, MS B283, Los Alamos National Laboratory, 
  Los Alamos, NM 87545, USA }

\emailAdd{yiannis@lanl.gov}
\emailAdd{ivitev@lanl.gov}

\abstract{
For heavy quarkonia of moderate energy, we generalize the relevant successful theory,  non-relativistic Quantum Chromodynamics (NRQCD), to include interactions in nuclear matter. The  new resulting theory, NRQCD with Glauber gluons,  provides for the first time a universal microscopic  description of the interaction of heavy quarkonia with a strongly interacting medium, consistently applicable to a range of phases, such as cold nuclear matter, dense hadron gas, and quark-gluon plasma.  The effective field theory  we present in this work is derived from first principles and  is an important  step forward in understanding the  common trends in proton-nucleus  and  nucleus-nucleus  data on quarkonium suppression.  
}

\begin{document}

\maketitle


\section{Introduction}

It is widely believed today that novel phases of nuclear matter, such as the  quark-gluon plasma (QGP) and a hot, dense gas of hadrons, are integral and important parts of the evolution of the  early universe. These extreme environments  are inaccessible to direct observation, but can be recreated in the laboratory  by colliding  heavy nuclei at relativistic energies. One of the main goals of nuclear physics is to accurately determine the properties of these new states of matter~\cite{BraunMunzinger:2007zz}. Since their lifetimes are very short, of order $10^{-23}$ s, one must use the produced particles themselves to probe the QGP and the hadron gas.  Quarkonia have emerged as premier diagnostics  of the QGP.  It was predicted that, when immersed in the plasma characterized by very high temperature, the color interaction between the heavy quarks will be screened  and quarkonia will dissociate~\cite{Matsui:1986dk}. Excited, weakly-bound states are expected  to melt away first, ground  tightly-bound states are expected to  melt away last, provide a way to determine the plasma temperature~\cite{Mocsy:2007jz}.

In the past decade phenomenological studies of quarkonia have evolved significantly to include effects that range from heavy quark recombination to dissociation through collisional interactions of $J/\psi$ and $\Upsilon$ states propagating through the QGP~\cite{Krouppa:2015yoa,Hoelck:2016tqf,Du:2017qkv,Aronson:2017ymv,Jamal:2018mog}.  The physics input in such calculations comes from the hard thermal loop calculations of the real and imaginary parts of the heavy quark-antiquark potentials~\cite{Laine:2006ns,Brambilla:2008cx}, lattice QCD calculations~\cite{Burnier:2015tda},  a $T-$matrix approach~\cite{Riek:2010fk} to obtain  interaction and  decay rates of thermal states, and lightcone wavefunction approach to obtain the dissociation rate of  quarkonia from collisional and thermal effects~\cite{Sharma:2012dy}. The evolution of the quarkonium system has been described by rate equations~\cite{Sharma:2012dy,Ferreiro:2014bia}, stochastic equations~\cite{Akamatsu:2011se,Brambilla:2016wgg,Kajimoto:2017rel} such as the Lindblad equation,  and the Boltzmann equation~\cite{Yao:2018nmy}.  Those studies has focused almost exclusively on quarkonia in a thermal QGP medium.

In spite of the advances described above, a fully coherent theoretical picture of  quarkonium production at the Relativistic Heavy Ion Collider (RHIC) and the Large Hadron Collider (LHC)  has not yet emerged.  In proton-nucleus (p+A) collisions, where QGP is much less likely  to be formed, attenuation similar to the one seen in nucleus-nucleus (A+A) reactions is still observed, albeit of smaller magnitude.  Even in high multiplicity  proton-proton (p+p) collisions there is evidence for $\Upsilon(2S)$ disappearance as a  function of the hadronic activity (${\rm N}_{\rm tracks}$) in the  event.   Specifically,  the relative suppression of the excited  versus  ground bottomonium states  $\Upsilon(2S)/\Upsilon(1S)$  as a function of the number of charged particle tracks,   shows  the same dissociation trend for  high-multiplicity proton-proton, proton-lead, and lead-lead reactions at the LHC~\cite{Chatrchyan:2013nza}. This experimental finding  has not yet found satisfactory theoretical expectations. It was argued  very recently that quarkonium  dissociation by co-movers  might be responsible for those trends~\cite{Ferreiro:2018wbd}. Differential  $\psi', \, \chi_c$ and $\Upsilon$ suppression was also established at RHIC~\cite{Adare:2013ezl,Adamczyk:2013poh} in d+Au reactions.  Upcoming experimental detector upgrades at RHIC and luminosity upgrades at the LHC will allow extensive studies of $J/\psi$ and $\Upsilon$ states with improved precision in high-multiplicity hadronic and nuclear collisions.   There is an opportunity to  further develop {\em microscopic} QCD approaches that describe this quarkonium physics in nuclear matter and that will facilitate  the quantitative determination  of the transport properties of the QGP and the hadron gas.

With this motivation, we first notice that calculations of heavy quarkonium production encounter hierarchies of momentum and mass scales, which is precisely where effective filed theories (EFTs) excel in reducing theoretical uncertainties and improving computational accuracy~\cite{Almeida:2014uva}.  Usually the scales one encounters are $p_T$, $m_Q$, $m_Q \ve$, $m_Q \ve^2$, and $\Lambda_{\text{QCD}}$, where $p_T$ is the quarkonium transverse momentum, $m_Q$ the heavy quark mass, and $\ve$ the heavy quark-antiquark pair relative velocity in the quarkonium rest frame. For moderate and high transverse momentum $p_T \gtrsim 2m_Q$ the established and most successful theory that describes quarkonium production and decays is non-relativistic QCD (NRQCD)~\cite{Bodwin:1994jh}.  Many recent theoretical studies take full advantage of the EFT capabilities  to significantly boost the theoretical precision of $J/\psi$ and $\Upsilon$  analyses and propose modern observables~\cite{Lansberg:2019adr} that can probe the quarkonium production mechanisms. Most of those studies focus their efforts on  quarkonium states in the  high energy ($E \gg m_{Q{\overline Q}}$) region, where theoretical advances  are now possible based  upon NRQCD, SCET~\cite{Bauer:2000ew,Bauer:2000yr,Bauer:2001ct,Bauer:2001yt}, and the picture of parton fragmentation~\cite{Baumgart:2014upa,Bain:2016rrv}.

The challenge that we face is to develop a microscopic theory of quarkonia applicable to different phases of nuclear matter in p+A  and A+A reactions. We approach this challenge from the effective field theory point of view.  The distinct advantage of an EFT approach is that it can provide a model-independent description of the universal  physics of energetic particle production in the background of a QCD medium.  This universal description can be applied equally well to the QGP or to a hadron gas, with model dependence entering only in the choice of the medium. In the past several years there were important developments in applying an EFT approach to describe particle production in the presence of strongly interacting matter. Particularly relevant to this work is the formulation and application of an effective theory of QCD, soft collinear effective theory with Glauber gluons (SCET${_{\rm G}}$)~\cite{Idilbi:2008vm,Ovanesyan:2011xy} for light particles ($\pi^{0,\pm}$, $K^\pm$, $\cdots$).   It was also demonstrated  that rigorous treatment of heavy flavor in matter is possible by constructing the necessary extension of SCET$_{\rm G}$ to nonzero quark masses, giving us the applicable  theory for energetic mesons containing a single heavy quark~\cite{Kang:2016ofv}.  SCET${_{\rm G}}$ allowed us for the first time, to overcome known limitations of traditional phenomenological approaches, use the same computational techniques  in high energy and heavy ion physics, and increase the accuracy and quantify the theoretical uncertainties  in the  calculations of light particle~\cite{Kang:2014xsa, Chien:2015vja} and heavy meson~\cite{Kang:2016ofv} production in A+A reactions.

As is the case in the vacuum, production of quarkonia in nuclear matter remains a multi-scale problem. For this reason, we identify the EFT approach the correct way to attack it. In this paper we demonstrate how one can generalize NRQCD to incorporate interactions of the non-relativistic heavy quarks with the medium. This is achieved through incorporating the  Glauber and Coulomb gluon exchanges of the heavy quarks with three different sources: collinear, soft, and static. We believe this version of NRQCD  will facilitate a much more  robust and accurate theoretical analysis of  the wealth of quarkonium measurements
in dense QCD matter.

The outline of this paper is as follows: In Section~\ref{sec:energy_loss},  after a brief overview of NRQCD, we explore the applicability of  the well-established  energy loss approach to quarkonia. We take the leading power factorization limit,  where a quarkonium state is produced thought the fragmentation process from a parton that undergoes energy loss in matter and demonstrate that the predicted  magnitude and hierarchy of suppression for ground and excited charmonium states is not compatible with  the experimental data. With this in mind, we,  consider the  propagation of the quarkonium state itself in QCD matter in Section~\ref{sec:scalings}. The possible off-shell gluon exchanges between the heavy quark/antiquark and the medium are discussed for several sources of scattering and we identify two relevant modes that mediate the interaction: Coulomb and Glauber gluons. In the following Section~\ref{sec:F_rules}, we give the Lagrangian and derive the Feynman rules for such exchanges. Finally, we conclude in Section~\ref{sec:conclusions}.  We discuss how a self-consistent background field approach to quarkonium propagation in matter can be
formulated in Appendix~A.


\section{Energy loss approach within the NRQCD formalism}
\label{sec:energy_loss}
Before we proceed to  the formulation of a generic effective theory of quarkonium production in matter, we have to explore whether medium-induced radiative processes might contribute significantly to the  modification of quarkonium cross sections in reactions with nuclei. It was suggested~\cite{Spousta:2016agr,Arleo:2017ntr}  that such effects can reduce the cross section of high transverse momentum $J/\psi$ production at the LHC~\cite{Khachatryan:2016ypw,Aaboud:2018quy}.

After we give a brief review of vNRQCD we proceed by describing  the leading power factorization of NRQCD for quarkonium production and introduce the quarkonium fragmentation functions within the NRQCD framework. We then apply energy loss effects to obtain quarkonium production rates in medium.


\subsection{Non-relativistic QCD: a brief overview}
In the quarkonium rest frame, the heavy quark and antiquark have small relative velocity, ($\lambda^2\sim 0.1$ for bottomonium  and $\lambda^2 \sim 0.3$ for charmonium). Therefore, NRQCD, which is an effective field theory that describes Quantum Chromodynamics in the non-relativistic limit, provides the correct theoretical framework for studying their interactions. 

There are three important scales that appear when studying the dynamics of non-relativistic heavy quarks: the mass of the heavy quark, $m$, the size of their momentum in the quarkonium rest frame, $m \ve$, and their kinetic energy, $m \ve^2$. The distance $r\sim 1/(m\ve)$ gives an estimate on the size of the quarkonium state and the separation between the heavy quark-antiquark pair. The non-relativistic kinetic energy $\Delta E \sim m \ve^2$ is of the same order as the energy splittings of radial excitations. We refer to $m \ve$ and $m \ve^2$ as the soft and ultra-soft scales respectively. Correspondingly, gluons that have all of their four-momentum components scaling as $m \ve$ and $m \ve^2$ are called soft and ultra-soft gluons. While the ultra-soft scale is well within the non-perturbative regime the soft scale is about 1.5 GeV for both bottomonium and charmonium. 

The effective theory of vNRQCD is a version non-relativistic QCD introduced in Ref.~\cite{Luke:1999kz} and recently formulated in a manifestly gauge invariant form in Ref.~\cite{Rothstein:2018dzq}. What we find appealing about this version of NRQCD is the clear distinction of soft and ultra-soft degrees of freedom and the use of  label-momentum notation. Both of those aspects are essential for the purposes of our work. We work in the limit where the measurement is sensitive to the kinematics of the heavy quark-antiquark pair (in the quarkonium rest frame) and therefore is critical we can separate the various infrared degrees of freedom. Using the four-vector $v^{\mu} = (1,\bmat{0})$, the four-momenta of the heavy quark, $p$, can be written as follows,
\begin{equation}
  \label{eq:momenta-deco}
  p^{\mu} = m v^{\mu} + r^{\mu} \;,
\end{equation}
where $r_0$ is the kinetic energy and $\bmat{r}$ is the three momentum of the heavy quark. Since the heavy quarks we  consider are on-shell, i.e. $p^2=m^2$, then in the non-relativistic limit where the three momentum is small compared to the mass,  $ \vert \bmat{r} \vert \sim \lambda m$, with $\lambda \ll 1$  we have
\begin{align}
  p^2 = m^2 + m r_0 + (r_0)^2 - \bmat{r}^2 = m^2 \;,
\end{align}
which has solution only if $ r^{\mu} \sim (\lambda^2, \bmat{\lambda}) $. In the presence of both soft and ultra-soft modes, it is important to decompose the small momentum component in its soft (label) and ultra-soft (residual) parts,
\begin{equation}
  \label{eq:deco2}
  p^{\mu} = m v^{\mu} + r_{us}^{\mu}  + r_{s}^{\mu}\;, 
\end{equation}
where $r_{us}^{\mu} \sim (\lambda^2,\lambda^2,\lambda^2,\lambda^2),$ and $ r_{s}^{\mu} \sim  (0,\lambda^1,\lambda^1,\lambda^1) $.
Then the connection with the convention in Eq.~(\ref{eq:momenta-deco}) can be made with the replacement,
\begin{align}
  \label{eq:deco1-deco2}
  r_0 &=  r_{0,us}\;, & \bmat{r} &=  \bmat{r}_{s} +\bmat{r}_{us} \;.
\end{align}

The QCD heavy quark field ($\Psi$) can then be decomposed in the vNRQCD heavy quark field ($\psi_{\bmat{\ell}}(x)$) as follows,
\begin{equation}
  \Psi(x) = \sum_\bmat{\ell} e^{-i \bmat{\ell} \cdot \bmat{x}} \psi_{\bmat{\ell}}(x)  \, , 
\end{equation}
where $\ell$ are the label components of the heavy quark momentum and $x$ is the coordinate space conjugate of the residual components. The soft ($A_{\ell}^{\mu}$) and ultra-soft ($A_{us}^{\mu}$) gluon fields have momenta which scale (all four components) as soft ($\sim m\ve$) or ultra-soft ($\sim m\ve^2$) respectively.

The Lagrangian of the EFT can then be written in terms of those fields in the following form~\cite{Luke:1999kz, Rothstein:2018dzq},
\begin{multline}
  \label{eq:L-vNRQCD}
  \mathcal{L}_{\text{vNRQCD}} = \sum_{\bmat{p}} \psi^{\dag}_{\bmat{p}} \lp iD^0 - \frac{(\bmat{\mathcal{P}}-i\bmat{D})^2}{2m} \rp \psi_{\bmat{p}} +\mathcal{L}^{(2)} + (\psi \to \chi, T \to \bar{T}) \\
  +  \mathcal{L}_s(\phi,\bar{\phi},A_q^{\mu})  +  \mathcal{L}^{V}(\psi,\chi,A_q^{\mu}) \;,
\end{multline}
where $\psi$ denotes the heavy quark field and $\chi$ the corresponding antiquark. The Lagrangian terms $\mathcal{L}^{(2)}$ are higher order terms, $\mathcal{L}_s$ is the soft gluon and ghost part of the Lagrangian, and $ \mathcal{L}^{V}$ contains the potential terms which have the following generic structure,
\begin{align}
  \text{Double soft gluon emissions:}&\;\; \sum_{\bmat{p},\bmat{p}',\ell,\ell'}\psi^{\dag}_{\bmat{p}} \lp A^{\mu}_{\ell} \; A^{\nu}_{\ell'} \rp   \psi_{\bmat{p}'} U_{\mu \nu} (\bmat{p},\bmat{p}',\ell,\ell') \;, \nn\\
   \text{Interactions with soft fermions:}&\;\; \sum_{\bmat{p},\bmat{p}',\ell,\ell'} \lp \psi^{\dag}_{\bmat{p}} T^{A}  \psi_{\bmat{p}'} \rp \lp \bar{\phi}_\ell T^A \gamma^{\mu} \phi_{\ell'}\rp Z_{\mu} (\bmat{p},\bmat{p}',\ell,\ell') \;, \nn\\ 
  \text{Heavy quark-antiquark  potential:}&\;\; \sum_{\bmat{p},\bmat{p}'}\lp \psi^{\dag}_{\bmat{p}} T^A   \psi_{\bmat{p}'} \rp  \lp \chi^{\dag}_{-\bmat{p}} \bar{T}^A \chi_{-\bmat{p}'} \rp V (\bmat{p},\bmat{p}') \nn\;.
\end{align}
where $U_{\mu,\nu}$, $Z_{\mu}$, and $V$ are functions of the momenta of the field included in the corresponding interactions. The soft fermion fields, $\bar{\phi}_{\ell}$, acting on the vacuum creates a light quark with soft momenta, $\ell^{\mu} \sim (\lambda,\lambda,\lambda,\lambda)$, and similarly $\phi_{\ell}$ for the antiquark. The Lagrangian that describes the interaction of soft fermions with soft gluons is identical to QCD, see Ref~\cite{Rothstein:2018dzq}.
The label momentum operator~\cite{Bauer:2001ct}, $\mathcal{P}^{\mu} = (\mathcal{P}^0, -\bmat{\mathcal{P}})$, is defined such that it projects only onto the label momentum space,
\begin{align}
  \mathcal{P}^{\mu}  \psi_{\bmat{\ell}}(x) &= \ell^{\mu}  \psi_{\bmat{\ell}}(x)\;, &  \mathcal{P}^{\mu} A_{\ell}^{\nu} &= \ell^{\mu} A_{\ell}^{\nu} \;, &  \mathcal{P}^{\mu} A_{us}^{\nu} &= 0 \; .
\end{align}
and the covariant derivative is  $iD^{\mu} \equiv i\partial^{\mu} - g A_{us}^{\mu}(x)$.

In collider physics, quarkonium production is studied within the NRQCD factorization conjecture, based on which  the cross section is written as a sum of products of short distance matching coefficients and the corresponding long distance matrix elements (LDMEs)
\begin{equation}
  \label{eq:nrqcd-fact}
  d\sigma_{ij \to \mathcal{Q} +X}(p_T) = \sum_n d\sigma_{ij \to Q\bar{Q}[n] + X'}(p_T) \langle \mathcal{O}^{\mathcal{Q}}(n) \rangle  \;.
  \end{equation}
The short distance coefficients (SDCs), $d\sigma_{ij \to Q\bar{Q}[n] + X'}$, describe the production of the $Q\bar{Q}[n]$ pair in a particular angular momentum and color configuration, $n=\Q{2S+1}{L}{J}{c}$. In the case of hadronic initial states, SDCs are expressed as a convolution of the partonic cross section and the collinear PDFs. The partonic cross section is then calculated in the matching of NRQCD onto QCD as an expansion in the strong coupling constant~\cite{Cho:1995vh, Cho:1995ce, Petrelli:1997ge,Braaten:1994kd, Ma:1995vi, Braaten:1996rp, Braaten:1994vv, Bodwin:2015iua}. In contrast, the LDMEs, $\langle \mathcal{O}^{\mathcal{Q}}(n) \rangle$, describe the decay of the $Q\bar{Q}[n]$ pair into the final color-singlet quarkonium state, $\mathcal{Q}$, through soft and ultra-soft gluon emissions. LDMEs are universal and fundamentally non-perturbative objects, and need to be extracted from experiment~\cite{Butenschoen:2011yh, Butenschoen:2012qr, Chao:2012iv, Bodwin:2014gia,  Bodwin:2015iua}. Although in principle all possible intermediate $Q\bar{Q}[n]$ configurations contribute to the final quarkonium state, LDMEs scale with powers of $\ve$, thus, we can truncate the sum up to the desired accuracy.


\subsection{Quarkonium fragmentation functions}

In order to envision energy loss processes as contributors to the modification of quarkonium cross sections in QCD matter two conditions must be satisfied. 
First, quarkonium production must be expressed as fragmentation of partons into the various $J/\psi$ and $\Upsilon$ states. The energy of the hard
parton is then reduced through inelastic processes in matter prior to fragmentation.  Second, the process  of fragmentation of quarkonia must happen  at time scales larger than the size of the QCD medium, $\tau_{form} \geq L$. This  condition must  also be investigated phenomenologically in reactions with nuclei, as the simpler hadronic collisions do not give relevant constraints.

Fortunately, in the last decade a leading power (LP) factorization of NRQCD has been established~\cite{Kang:2014tta,Nayak:2005rw,Nayak:2005rt,Kang:2011mg,Kang:2011zza,Fleming:2012wy}  and is expected to hold at high transverse momenta ($p_T \gg m_{Q}$). In the large transverse momentum limit the NRQCD short distance coefficients suffer from logarithmic enhancements of the form $\alpha_s^{m} \ln^{n}(p_T/2m_{Q})$. These terms could spoil the perturbative expansion and, thus, resummation is necessary in order to make meaningful predictions. This is achieved through the LP factorization of NRQCD, where the cross section is now factorized into short distance matching coefficients (that describe the production and propagation of a parton $k$) and the so called NRQCD fragmentation functions,
\begin{equation}
  d\sigma_{ij \to \mathcal{Q} +X}(p_T) = \sum_n \int_{x_{\rm min}}^{1} \frac{dx}{x} d\sigma_{ij \to k + X'}\lp\frac{p_T}{x} , \mu \rp  D_{k/\mathcal{Q}}^{\;n}(x, \mu) \;.
\end{equation}
The dependence on the factorization scale, $\mu$,  of the factorized terms is exactly what allows for the resummation of large logarithms through the use of renormalization group techniques and, particularly, the DGLAP evolution for the fragmentation functions.  Comparison of the above equation with Eq.~(\ref{eq:nrqcd-fact}) immediately gives that  the NRQCD fragmentation functions can be written in terms of the same LDMEs that appear in the fixed order factorization and perturbatively calculable matching coefficients, 
\begin{equation}
   D_{k/\mathcal{Q}}^{\;n}(x, \mu) =  \frac{ \langle\mathcal{O}^{\mathcal{Q}}(n) \rangle} {m_c^{[n]}} \;  d_{k/n}(x,\mu) \;,  
  \end{equation}
where $[n] =0 $ for S-wave and  $[n]=2$ for P-wave quarkonia. The short distance coefficients, $d_{k/n}(x,\mu)$, are functions of the fraction, $x$, of the parton energy transferred to the quarkonium state. They describe the fragmentation of the initiating parton to an intermediate $Q\bar{Q}(\Q{2S+1}{L}{J}{1/8})$ pair. The LP factorization is expected to hold for $p_T \gg m_{Q}$ but the precise $p_T$  region of validity cannot be be determined analytically. However, phenomenological applications to charmonia have shown that it may hold to transverse momenta as low as $p_T =10$~GeV~\cite{Bodwin:2015iua}.

\begin{figure}[h!]
  \centerline{\includegraphics[width = 0.95 \textwidth]{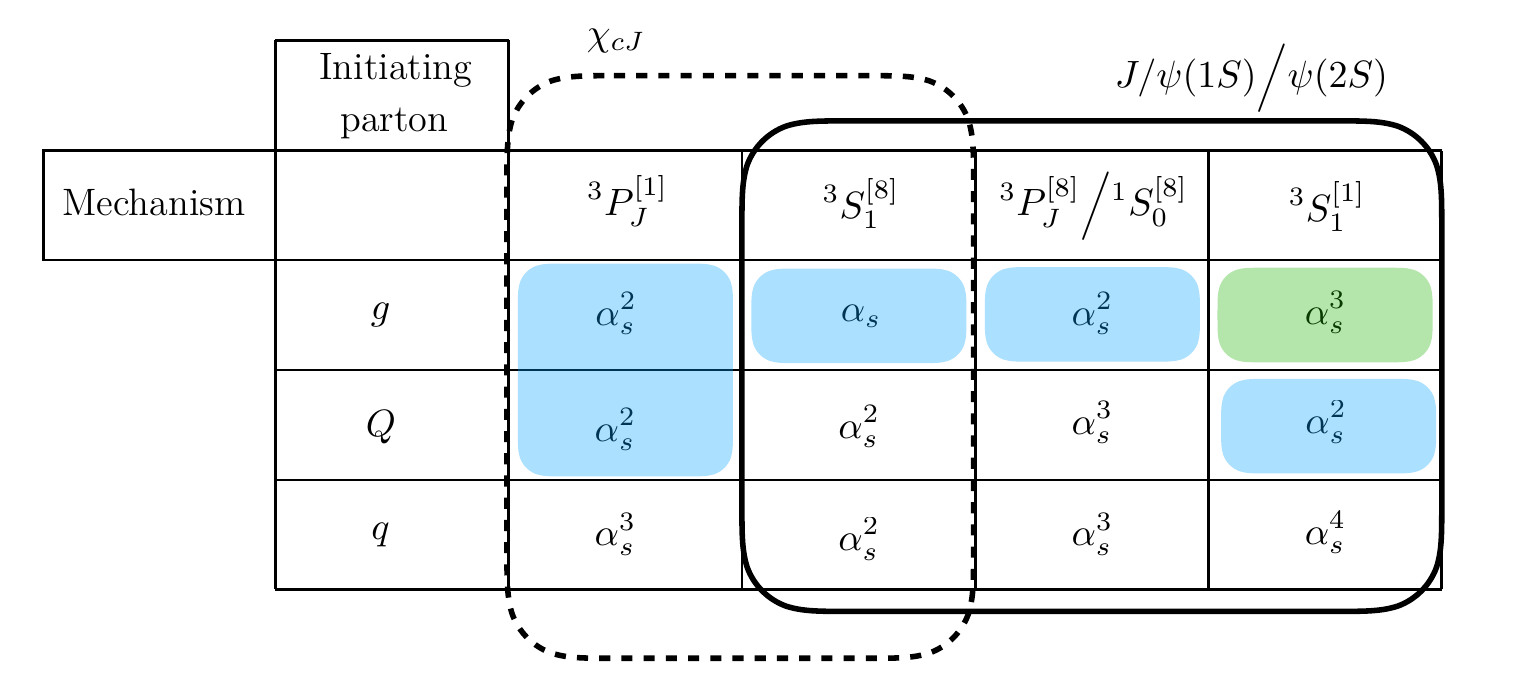}}
  \caption{Order in $\alpha_s$ for the leading fragmentation mechanisms for quarkonia. We include the light blue (leading per channel) and green shaded mechanisms.}
  \label{fig:channels}
\end{figure}

In this work we consider both the direct production and the feed-down from decays of excited quarkonium states. For $J/\psi$  the following feed-down contributions are implemented, 
\begin{align}
  \psi(2S):& \;\;\;\; \text{Br}\lb \psi(2S) \to J/\psi + X \rb = 61.4 \pm 0.6 \%  \;,  \nn \\
  \chi_{c1}:& \;\;\;\; \text{Br}\lb \chi_{c1} \to J/\psi + \gamma \rb = 34.3 \pm 1.0 \%  \;,  \nn \\
  \chi_{c2}:& \;\;\;\; \text{Br}\lb \chi_{c2} \to J/\psi + \gamma \rb = 19.0 \pm 0.5 \%  \;.  
\end{align}
For the direct fragmentation of a parton to $J/\psi$ and $\psi(2S)$ we consider the following intermediate $Q\bar{Q}$ states: $\Q{3}{S}{1}{8},\;\Q{1}{S}{0}{8},\;\Q{3}{P}{J}{8},$ and $\Q{3}{S}{1}{1}$. With exception of the $\Q{3}{S}{1}{1}$ channel, for each other channel we only conciser the leading in $\alpha_s$ contribution.  As a result, the  various channels will be evaluated at different order in the perturbative expansion. For the case $\Q{3}{S}{1}{1}$, where the leading mechanism is the heavy quark fragmentation, in addition we include the gluon channel due to the abundance of gluons in hadronic collisions. These contributions are summarized in Figure~\ref{fig:channels}.

The dominant production channels for the $\chi_{cJ}$ come from the intermediate $Q\bar{Q}[n] \to \chi_{cJ} $ states for which $n \in \{ \Q{3}{P}{J}{1}, \Q{3}{S}{1}{8}  \}$. For these mechanisms, we identify the  gluon and heavy quark initiating processes to be the most relevant, see Figure~\ref{fig:channels}. Therefore, the fragmentation functions we need for our analysis are:
\begin{align}
  D_{g/\chi_{cJ}}^{\Q{3}{S}{1}{8}} (z,2m_c) &= \langle\mathcal{O}^{\chi_{cJ}}(\Q{3}{S}{1}{8}) \rangle \;d_{g/\Q{3}{S}{1}{8}}(z,2 m_c)  \;,   \nn \\
  D_{g/\chi_{cJ}}^{\Q{3}{P}{J}{1}} (z,2m_c) &= \frac{ \langle\mathcal{O}^{\chi_{cJ}}(\Q{3}{P}{J}{1})  \rangle} {m_c^2}\;  d_{g/\Q{3}{P}{J}{1}}(z,2 m_c)  \;,  \nn \\
  D_{Q/\chi_{cJ}}^{\Q{3}{P}{J}{1}} (z,2m_c) &= \frac{ \langle\mathcal{O}^{\chi_{cJ}}(\Q{3}{P}{J}{1}) \rangle} {m_c^2} \;  d_{Q/\Q{3}{P}{J}{1}}(z,2m_c) \;,  
\end{align}
where the LDMEs in this equation are evaluated at scale $\mu_{\Lambda} = 2 m_c$. To evolve the fragmentation functions $ D_{i/\mathcal{Q}}^{[n]}$ to an arbitrary scale $\mu > 2 m_c$ we use the standard DGLAP evolution~\cite{Gribov:1972ri, Altarelli:1977zs, Dokshitzer:1977sg} at leading logarithmic (LL) accuracy. From Ref.~\cite{Braaten:1994kd} we have,
\begin{equation}
  d_{g/\Q{3}{P}{J}{1}}(z, 2m_c) = \frac{2 \alpha_s^2(2m_c)}{81}{m_c^3} \lb  z \mathcal{L}_0(1-z) + \frac{1}{(2J+1)} \lp Q_J \delta(1-z) +P_J (z) \rp  \rb \;.
  \end{equation}
For the same channel, the heavy quark short distance coefficients are given by:
\begin{equation}
   d_{Q/\Q{3}{P}{J}{1}}(z,2m_c) = \frac{\hat{D}_J(z,2m_c)}{m_c^3} \;,
\end{equation}
where $\hat{D}_J(z,2m_c)$ are given in Eq.~(3.3) of Ref.~\cite{Ma:1995vi}. For the octet production mechanism, $\Q{3}{S}{1}{8}$, also present in the case of $\psi(nS)$, we have (see Refs.~\cite{Braaten:1994vv, Braaten:1996rp}):
\begin{equation}
  d_{g/\Q{3}{S}{1}{8}}(z,2m_c) = \frac{\pi \alpha_s(2m_c)}{24 m_c^3} \delta(1-z) \;.
\end{equation}

Our analysis for the direct production of $J/\psi$ and $\psi(2S)$ follows  Ref.~\cite{Baumgart:2014upa}. All relevant fragmentation functions and the corresponding Mellin transforms are collected in the Appendix of Ref.~\cite{Baumgart:2014upa}. A comprehensive analysis and extraction of the non-perturbative LDMEs, consistent with LP factorization, is given  by Ref.~\cite{Bodwin:2015iua}. Throughout this paper we use their results for the values of the LDMEs.


\subsection{Medium-induced energy loss}

Let us now turn to the application of energy loss to quarkonium production. If a parton $c$ loses  momentum fraction $\epsilon$ during its  propagation 
in the medium to escape with momentum $p_{T_c}^{med}$, in the short distance hard process its momentum is given by 
$p_{T_c} = p_{T_c}^{med} / ( 1 - \epsilon )$.  This also gives rise  to an additional Jacobian factor $|d^2 p_{T_c}^{med} 
/ d^2 p_{T_c}| = ( 1- \epsilon)^2 $, similar to the $z^2$ factor in the factorization formula for  hadron production.  The cross section for hadron  production
and quarkonium production per elementary nucleon-nucleon ($NN$) collision in the leading power limit is then written down as
\begin{eqnarray}
\frac{1}{\langle N_{\rm coll.} \rangle } \frac{d \sigma^{h}_{med}}{dyd^2p_T} & = &  \sum_c \int_{z_{\rm min}}^1 dz   \int_{0}^1 d\epsilon \, P(\epsilon)  \,   \frac{d \sigma^{c} \left(\frac{p_T}{(1-\epsilon) z  } \right) }{dyd^2p_{T_c}}   \frac{1}{(1-\epsilon)^2z^2}    D_{h/_c}( z ) \; .
\label{hspectrum-quench} 
\end{eqnarray}
In Eq.~(\ref{hspectrum-quench}) we have omitted the renormalization and factorization scale dependences for brevity. $P(\epsilon)$ is the probability distribution for the hard parton $c$ to lose energy due to multiple gluon emission, 
$\frac{d \sigma^{c} (p_T) }{dyd^2p_{T_c}} $ is the hard partonic cross section, and  $ \langle N_{\rm coll.} \rangle $  is the average number of
binary nucleon-nucleon colliions.

\begin{figure}[h!]
\begin{center}
 \includegraphics[width =  0.65\textwidth]{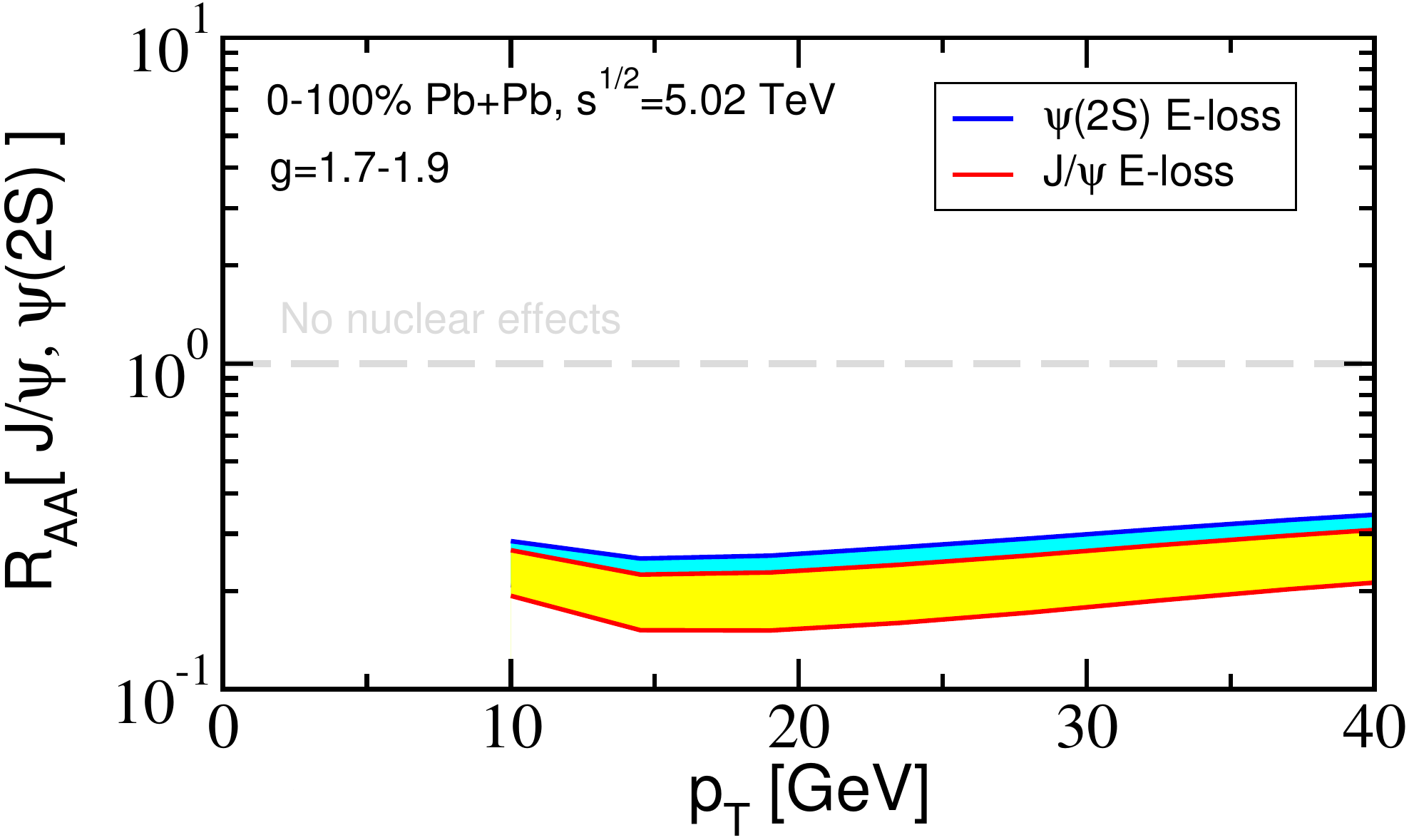} 
 \end{center}
  \caption{ Suppression of $J/\psi$ (yellow band) and  $\psi(2S)$ (cyan band)  cross sections in  minimum bias  lead-lead collisions at 
  $\sqrt{s_{NN}} = 5.02$~TeV. The band corresponds to a coupling between the parton and the medium $g=1.7 - 1.9$. }
  \label{fig:1S2Ssup}
\end{figure}

In the approximation  that the fluctuations of the average  number of medium-induced gluons are 
uncorrelated~\cite{Baier:2001yt,Gyulassy:2001nm}, the spectrum of the total radiative energy loss fraction due to multiple gluon emissions, 
$\epsilon=\sum_i \omega_i / E$,  can be expressed via a  Poisson expansion  $P(\epsilon,E)=\sum_{n=0}^\infty P_n(\epsilon,E)$,  
with $P_1(\epsilon,E)=e^{-\left\langle N^g \right\rangle} \rho(\epsilon,E)$.  We note that in our notation
  $\rho(x,E) $ is the medium-induced gluon spectrum 
\begin{equation}
 \rho(x,E)  \equiv \frac{dN^g}{dx}(x,E), \qquad    \int_{x_0}^{1-x_0}  \frac{dN^g}{dx}(x,E) = N^g(E)\;, 
\end{equation}
where  $x = \omega/E$ is the fraction of the energy of the parent parton taken by an individual gluon and 
$x_0=\Lambda_{\rm QCD}/2E$.   We keep explicitly the dependence on the parent parton energy but remark that
medium-induced gluon radiation also depends on the parton's  flavor and mass.
The terms of the Poisson series are generated iteratively as follows
\begin{eqnarray}
P_{n+1}(\epsilon,E)&=&  \frac{1}{n+1} \int_{x_0}^{1-x_0} dx_n \; \rho(x_n,E)
P_n(\epsilon-x_n,E) 
\nonumber \\[1ex]
&=&\frac{e^{-\left\langle{N^g(E)}\right\rangle}}{(n+1)!}\int dx_1\cdots 
dx_{n} \; \rho(x_1,E)\cdots\rho(x_{n},E)\rho(\epsilon-x_1-\cdots-x_{n},E)  \; . \quad
 \end{eqnarray}
We note that  in the presence of a medium radiation is attenuated at the typical Debye screening scale and the 
number of medium-induced gluons is finite.  Therefore, we have explicitly a finite $n=0$ no radiation 
contribution $P_0(\epsilon,E)= e^{-\left\langle {N^g(E)}\right\rangle }\delta(\epsilon)$.
The normalized Poisson distribution that enters Eq. (\ref{hspectrum-quench})  then gives
\begin{eqnarray}
\int_0^\infty d\epsilon \; P(\epsilon,E) \epsilon= \frac{\Delta E}{E}\;,   \qquad  \int_0^\infty d\epsilon \; P(\epsilon,E) =1 
\;.
\end{eqnarray}

\begin{figure}[h!]
\begin{center}
\includegraphics[width =  0.65\textwidth]{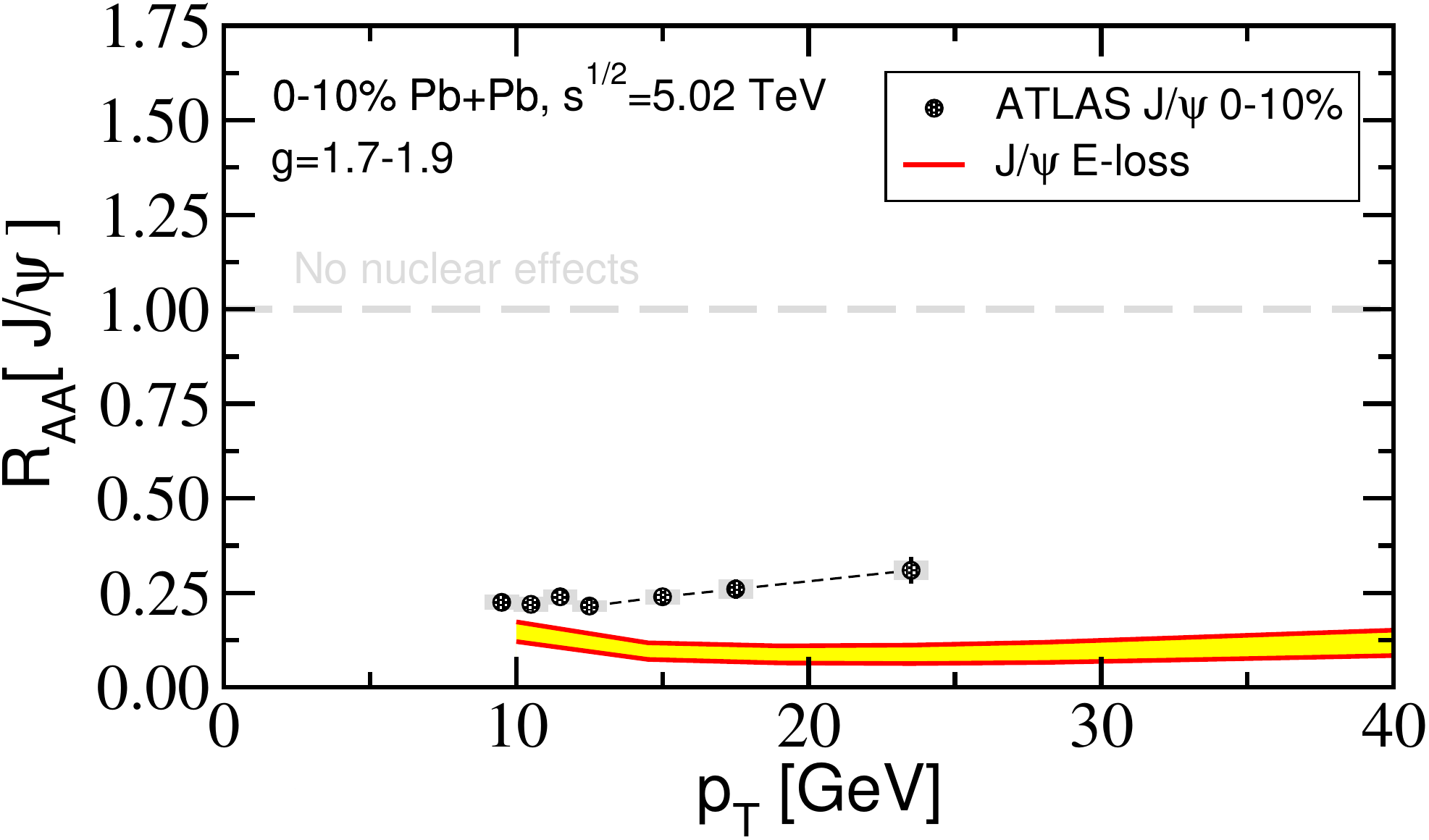} \\  
 \includegraphics[width =  0.65\textwidth]{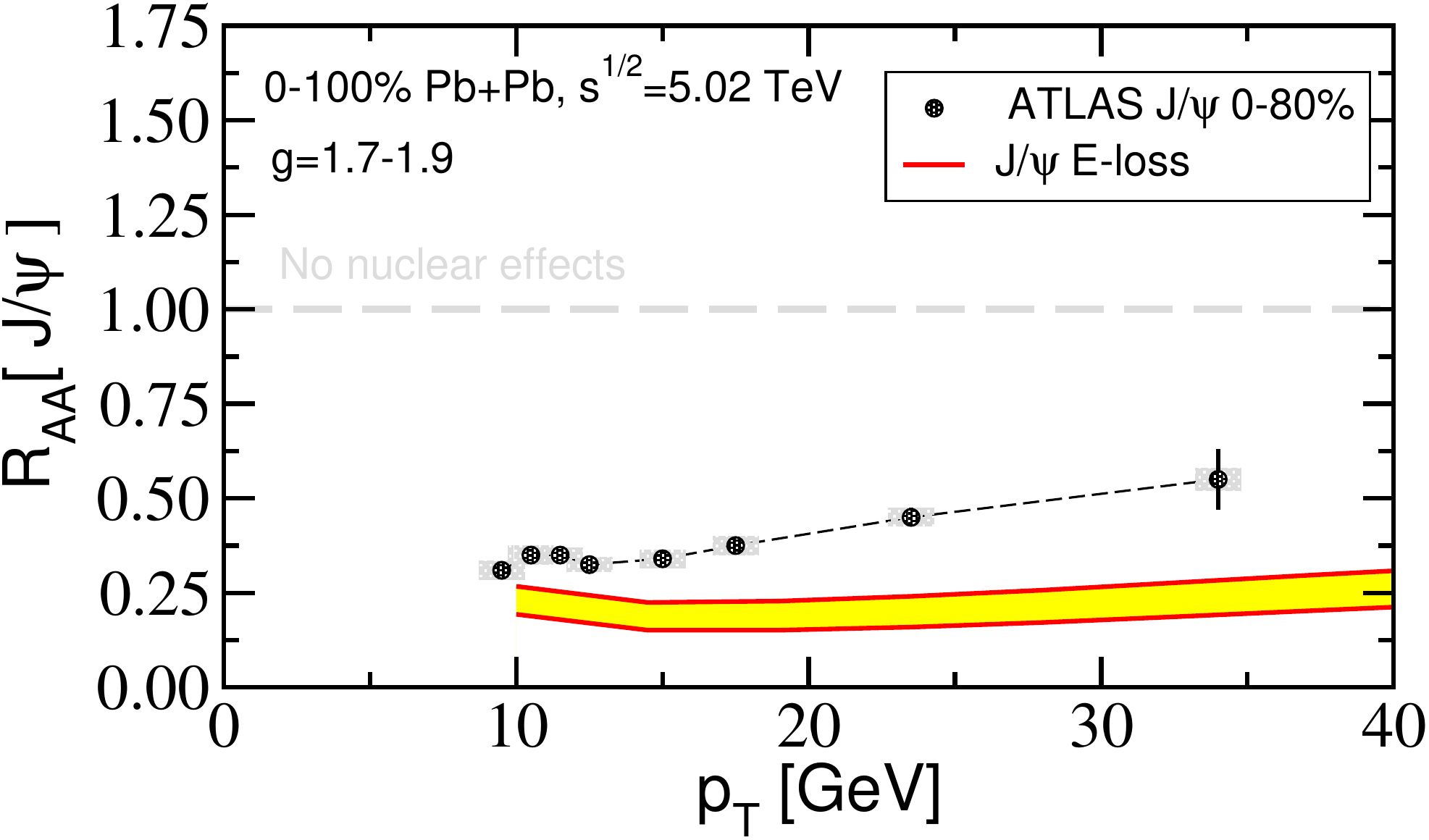} 
 \end{center}
  \caption{ Comparison of the  suppression of $J/\psi$ (yellow band) evaluated in an energy loss model with coupling between the parton and the medium $g=1.7 - 1.9$  to ATLAS data from $\sqrt{s_{NN}} = 5.02$~TeV Pb+Pb collisions at the LHC~\cite{Aaboud:2018quy}. Upper panel: comparison between theory and data in
  the most central 0-10\% collisions.   Lpper panel: comparison between theory and data in minimum bias collisions, the exact centrality class of ATLAS
  data is 0-80\%.   }
   \label{fig:1SsuppPTATLAS}
\end{figure}

Several formalisms have been developed in the literature to evaluate medium-induced gluon radiation~\cite{Zakharov:1997uu,Baier:1996kr,Gyulassy:2000er,Wang:2001ifa,Wang:200,Djordjevic:2003zk}. 
In this work, we use the soft gluon emission limit of the full in-medium splitting kernels~\cite{Ovanesyan:2011kn,Kang:2016ofv,Sievert:2019cwq}  
and evaluate them in a viscous 2+1 dimensional hydrodynamic  model of the  background medium~\cite{Shen:2014vra}.

We now turn to the  evaluation of the prompt $J/\psi$ and $\psi(2S)$ suppression in lead-lead (Pb+Pb) collisions at the LHC.  We calculate the 
partonic cross sections as in Ref.~\cite{Chien:2015vja}.  We chose the values of the coupling between the hard partons and the QCD medium that they 
propagate in  to be in the range $g=1.7 - 1.9$. These values are slightly smaller than the ones used in~\cite{Chien:2015vja}  and the difference can be 
traced to the different hydrodynamic models of the medium. Earlier works used ideal Bjorken expanding medium with purely gluonic degrees of freedom.
As we will show below, the suppression of quarkonia, especially the $J/\psi$, obtained in the energy loss framework is too large when compared to experimental 
measurements. Thus, if there is  an uncertainty in the choice of the coupling constant $g$, we must err on the side of smaller couplings. A larger coupling constant 
will produce an even larger discrepancy. Results are presented as the ratio of the cross sections in nucleus-nucleus (AA) collisions to the ones in   
nucleon-nucleon  collisions scaled with the number of binary nucleon-nucleon interactions
\begin{equation}
  R_{AA}= \frac{1}{\langle N_{\rm coll.} \rangle }
  \frac{d \sigma_{AA}^{\rm Quarkonia}/dy dp_T}{d \sigma_{pp}^{\rm Quarkonia}/dy dp_T}   \; .
\label{raa}  
\end{equation}

  In Figure~\ref{fig:1S2Ssup} we first show  the transverse momentum dependence of the of $J/\psi$ (yellow band) and  $\psi(2S)$ (cyan band) suppression.
  We use minimum bias Pb+Pb collisions at  $\sqrt{s_{NN}} = 5.02$~TeV for illustration and the suppression is calculated  as a sum over 
  centrality classes $i$ corresponding to mean impact parameters $b_i$  with weights   $W_i$~\cite{Aronson:2017ymv}
   \begin{eqnarray}
  R_{AA}^{\rm min.\ bias}(p_T)& =&
  \frac{\sum_i R_{AA}(\langle b_i \rangle) W_i}{\sum_i W_i} \quad {\rm where} \quad
  W_i =\int_{b_{i\, \min}}^{b_{i\, \max}}  N_{\rm coll.}(b)\,\pi \, b \,db  \, .
\label{cent}
\end{eqnarray}
We find that the theoretical calculation produces a rather flat  transverse momentum dependence of the quarkonium suppression factor  $R_{AA}$.
The magnitude of this suppression is large, a factor 3 to 5, and is very similar between the $J/\psi $ and  $\psi(2S)$ states. This is easy to understand, 
as in the parton energy loss picture the nuclear modification depends on the flavor and mass of the propagating parton, the fragmentation functions   
and the steepness of paticle spectra.  The ground and excited $J/\psi$ states have very similar partonic origin and fragmentation functions. The $\psi(2S)$
spectra are slightly harder than the ones for the  $J/\psi $  and this accounts for the slightly smaller suppression.

Comparison of theoretical calculations to ATLAS  experimental data on the  transverse momentum dependence of $J/\Psi$ attenuation from $\sqrt{s_{NN}} = 5.02$~TeV Pb+Pb collisions at the LHC~\cite{Aaboud:2018quy} is presented in  Figure~\ref{fig:1SsuppPTATLAS}. The top panel shows results for 0-10\% central collisions. As can be seen from the
figure, the data is not described by the theoretical predictions. Energy loss calculations overpredict the suppression of $J/\psi$ even in the lowest transverse
momentum bin around  $p_T \sim 10$~GeV.  At higher transverse momenta the discrepancy is as large as a factor of 3. The bottom panel of Figure~\ref{fig:1SsuppPTATLAS}
shows similar comparison but for minimum bias collisions (ATLAS measurements cover 0-80\% centrality). The same conclusion can be reached, i.e. the theoretical 
calculation predicts significantly the nuclear modification in comparison to the one measured  measured by the experiment.

\begin{figure}[h!]
\begin{center}
\includegraphics[width =  0.65\textwidth]{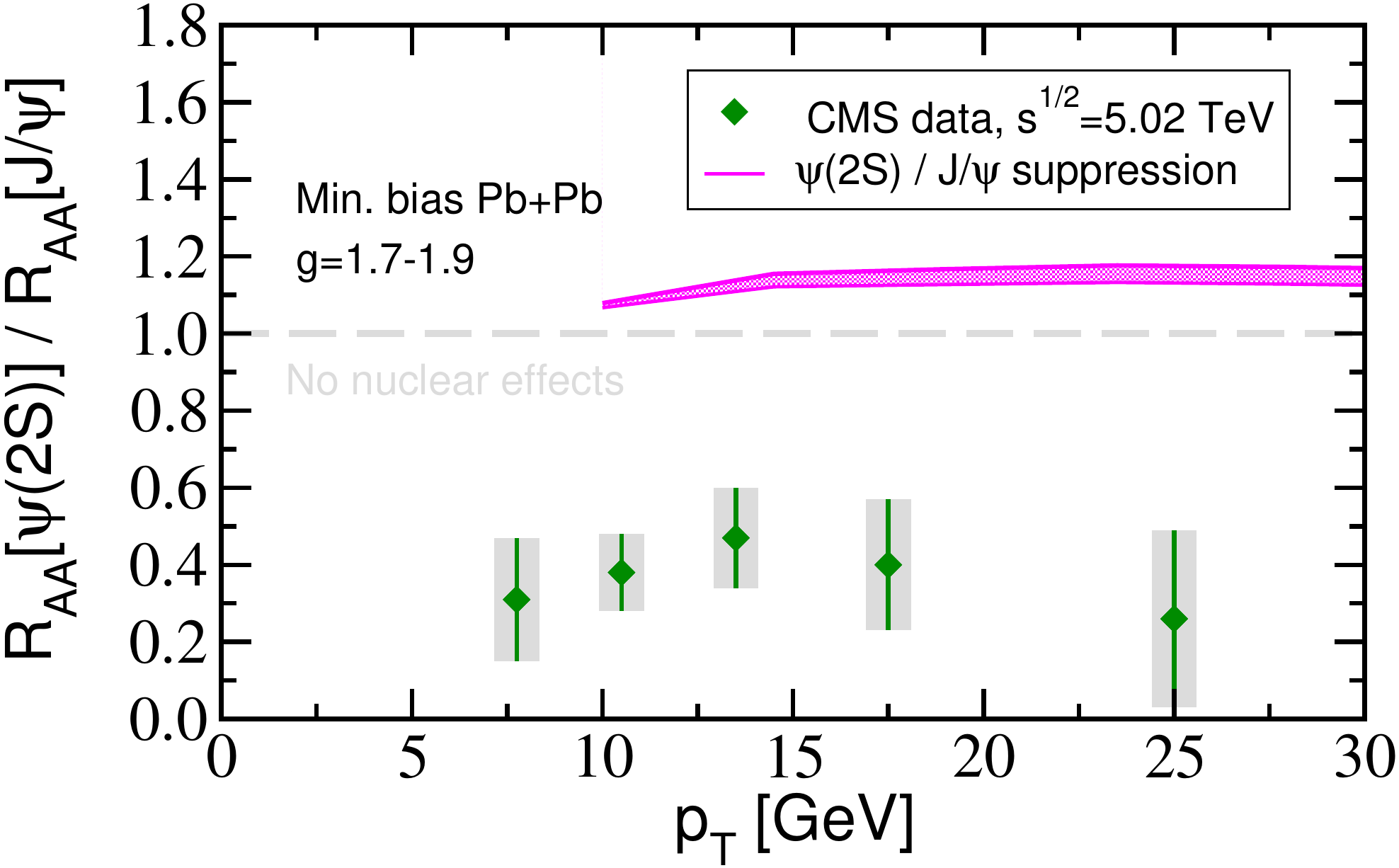} \\  
 \includegraphics[width =  0.65\textwidth]{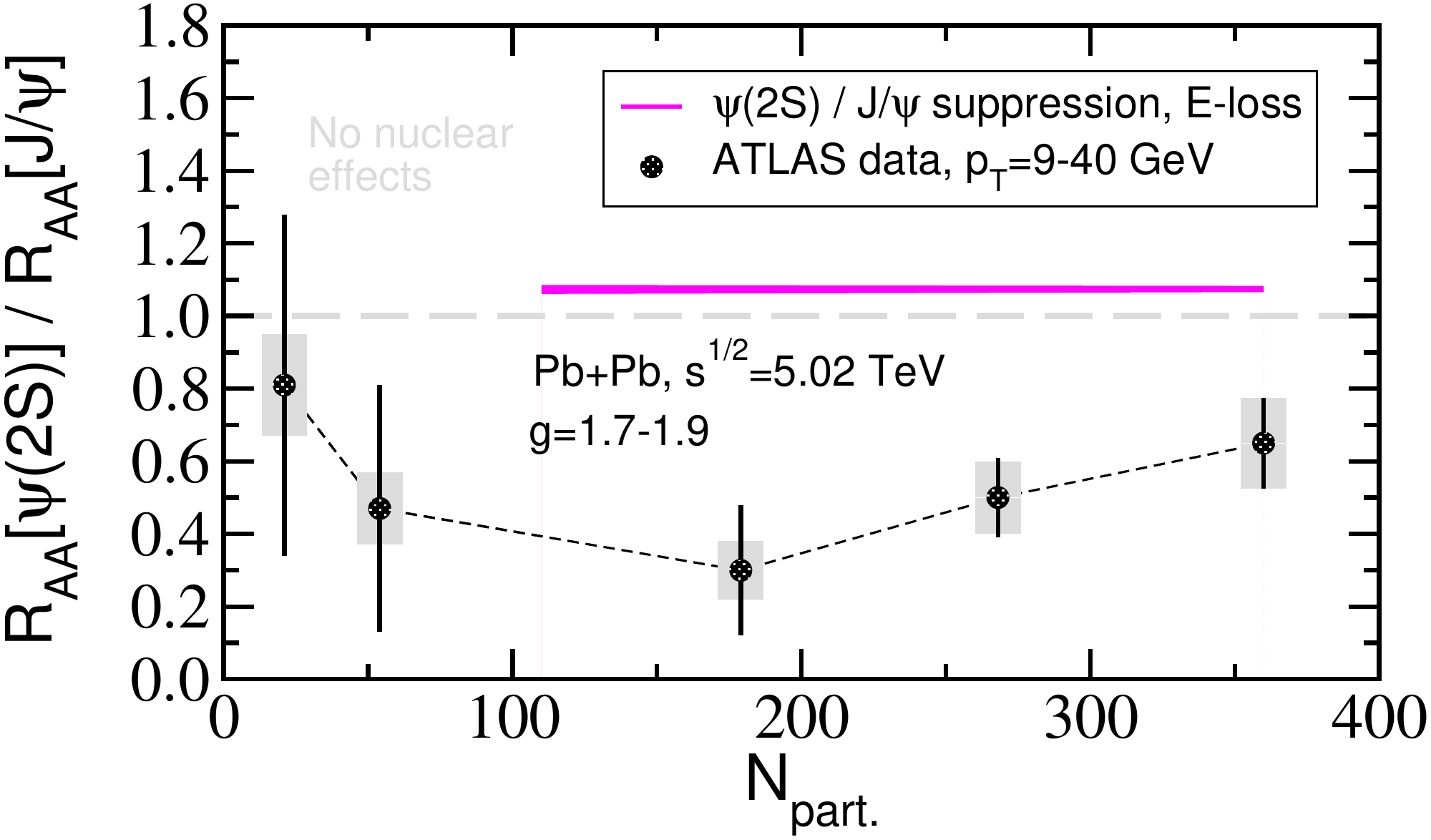} 
 \end{center}
  \caption{ The double ratio of $\psi(2S)$ to $J/\psi$ suppression (purple bands) as a measure of the relative significance of QCD matter effects on
  ground and excited states  is compared to energy loss model calculations. Upper panel:  comparison  between theory and CMS data~\cite{Khachatryan:2016ypw}  as a function of  transverse momentum $p_T$ for minimum bias collisions.  Lower panel:   comparison between theory and ATLAS data~\cite{Aaboud:2018quy} as a function of centrality  integrated in the  $p_T$
  region of 9-40~GeV.  } 
  \label{fig:2Sto1Sptcent}
\end{figure}

Next, we address the relative medium-induced suppression of $\psi(2S)$ to $J/\psi$ in matter in Figure~\ref{fig:2Sto1Sptcent}.  The purple bands correspond  to variation of the
coupling between the parton and the medium of $g=1.7-1.9$.  Since these are double ratios, the sensitivity to the variation of $g$ is significantly reduced. The upper panel of  Figure~\ref{fig:2Sto1Sptcent} shows the double nuclear modification ratio as a function of $p_T$ compared to CMS data~\cite{Khachatryan:2016ypw}.  Theory and experimental measurements 
are for minimum bias collisions and are clearly very different. The energy loss model predicts slightly smaller suppression for the  $\psi(2S)$ state when compared to $J/\psi$  and the double ratio is 10-20\% above unity. In contrast,  experimental results show that the suppression of the weakly bound   $\psi(2S)$ is 2 to 3 times larger than that of
$J/\psi$. It is clear that the energy loss model  is incompatible with the  hierarchy of excited to ground state suppression of quarkonia in matter.   The bottom panel of  Figure~\ref{fig:2Sto1Sptcent} shows the same ratio as a function of the number of participants $N_{\rm part.}$ and the transverse momenta are integrated in the range of 9-40~GeV. 
Similar conclusion about the tension between data and the theoretical model calculations can be reached, which is inherent to the model and cannot be resolved by 
varying the coupling between the partons that fragment into quarkonia and the medium.

In summary, in this section we demonstrated that in the currently accessible transverse  momentum range of up to $\sim 50$~GeV for quarkonium measurements in 
heavy ion collisions, the energy loss approach  combined with leading power factorization is not compatible with existing experimental data from the LHC. The tensions 
are  both in the overall magnitude of $J/\psi$ suppression  and in the  relative suppression of the    $\psi(2S)$  to the ground $J/\psi$.
This implies that the quarkonium states coexist with the medium and  motivates us to pursue the formulation a general theory for quarkonium interactions with nuclear matter.


\section{Toward a formulation of NRQCD$_{\rm G}$: the Glauber and Coulomb regions}
\label{sec:scalings}

The main goal of this work is to  devise a framework where quarkonia propagate in a variety of strongly-interacting media, such as cold nuclear matter, QGP, or a hadron gas. We are interested in the regime where matter itself might be non-perturbative, but the interaction with its  quasiparticles is mediated by gluon fields and can be described by perturbation theory. Such approach has proven to be extremely successful in constructing  theories of light flavor, heavy flavor, and jet production in heavy ion collisions.

When an energetic particle propagates in matter, the interaction with the quasiparticles of the medium is  typically mediated  by $t-$channel exchanges of off-shell gluons, called Glauber gluons.  We will, thus, call the new effective theory NRQCD with Glauber gluons, or NRQCD$_{\rm G}$. We have noticed in the past~\cite{Ovanesyan:2011xy}  that when the  sources  of interaction do not have large momentum component,  the exchange gluon field's momentum can scale as soft. Here, we call them Coulomb gluons and treat this limit explicitly. The Lagrangian of NRQCD$_{\rm G}$ is constructed by adding to the vNRQCD Lagrangian the additional terms that include the interactions with quark and gluon sources through (virtual) Glauber/Coulomb gluons exchanges. We may then write,
\begin{equation}
  \mathcal{L}_{\text{NRQCD}_{\rm G}} = \mathcal{L}_{\text{vNRQCD}} + \mathcal{L}_{Q-G/C} (\psi,A_{G/C}^{\mu,a}) + \mathcal{L}_{\bar{Q}-G/C} (\chi,A_{G/C}^{\mu,a})\;,
\end{equation}
where the effective fields $A_{G/C}^{\mu,a}$ incorporate the information about the source fields. In order to extract the form and perform the power-counting of the terms in  $\mathcal{L}_{Q-G/C} (\psi,A_{G/C}^{\mu,a})$ we will follow three different approaches: 
\begin{enumerate}
\item Perform a shift in the gluon field in the NRQCD Lagrangian ($A^{\mu}_{us} \to A^{\mu}_{us} +A^{\mu}_{G/C}$)  and then perform the power-counting established in Table~\ref{tb:scaling-A} to keep the leading contributions. This approach is also known as the background field method.
\item A hybrid method, where from the full QCD diagrams for single effective Glauber/Coulomb gluon insertion, and after performing the corresponding power-counting, one can read the Feynman rules for the relevant interactions. 
\item A matching method  where we expand in the power-counting parameter, $\lambda$, the full QCD diagrams describing the interactions of an incoming heavy quark and a light quark or a gluon. To get the NRQCD$_{\rm G}$ Lagrangian, we then keep the leading and subleading contributions and focus on the dominant contributions in forward scattering limit.  In contrast to the  hybrid method, here we also derive the tree level expressions of the effective fields in terms of the QCD ingredients.  
\end{enumerate}
The first two methods do not directly involve the source fields, since this information is compressed in the effective fields, $A_{G/C}^{\mu,a}$. We show that the background field method, naively applied in the vNRQCD Lagrangian, yields an ambiguous result. In Appendix~\ref{app:last} we discuss how to properly implement this method in agreement with the other two methods. The fact that all three approaches then give the same Lagrangian is a non-trivial test of our derivation.

We now consider the scaling of the gluon momenta, $q^{\mu}_{G/C}$, for the Glauber and Coulomb regions and the corresponding scaling of the effective gluon fields, $A^{\mu}_{G/C}$. This is done for three types of sources: collinear, soft, and static. We will use the four-component notation $(p^{0},p^{1},p^{2},p^{3})$ rather than the light-cone coordinates, $(p^{+},p^{-},p^{\perp})$, since this more compatible with the NRQCD formalism. We use $\bmat{n} = (0,0,1)$ as the direction of motion of the collinear source.  

\begin{figure}[h!]
  \centerline{\includegraphics[width =  \textwidth]{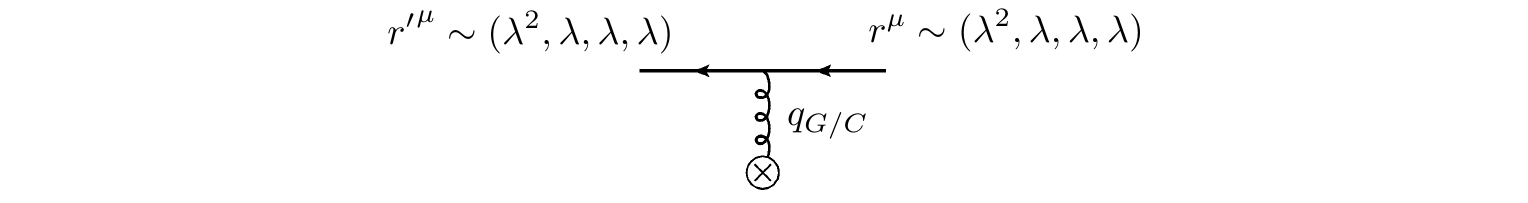}}
  \caption{A characteristic single Glauber/Coulomb gluon insertion vertex from the Lagrangian $\mathcal{L}_{Q-G/C}$, where the incoming quark caries momentum $p^{\mu} = m v^{\mu} + r^{\mu}$ and the outgoing ${p'}^{\mu} = m v^{\mu} + {r'}^{\mu}$.}
  \label{fig:heavy}
\end{figure}

Note that, for any gluon interacting with the vNRQCD heavy quark, we require  $q^{0}_{G/C} \sim \lambda^2$  and $q^i_{G/C} \lesssim \lambda$ such that the heavy quark momenta, both on the left and right of the insertion, scale as $(\lambda^2,\bmat{\lambda})$, as illustrated in Figure~\ref{fig:heavy}. If all of the three-momenta components scale as $\lambda$, i.e. $q_C^{\mu} \sim (\lambda^2,\bmat{\lambda})$ then this corresponds to Coulomb (or potential) gluons. The exchange of such modes between the heavy quarks and soft particle has already been investigated up to next-to-next-leading order in the non-relativistic limit in vNRQCD~\cite{Luke:1999kz, Rothstein:2018dzq}. We compare our derivations with theirs in Section~\ref{sec:comp}. On the other hand, collinear particles cannot interact with the heavy quarks through the exchange of Coulomb gluons since this will push the collinear particles away from their canonical angular scaling. The relevant mode here is the Glauber gluons, which scale as $q_G^{\mu} \sim (\lambda^2,\lambda,\lambda,\lambda^2)$. We will, therefore, consider Coulomb gluons for the interaction of the heavy quarks with soft and static modes and Glauber gluons for the interactions with collinear modes:
\begin{align}
  \text{for static and soft sources:}\;\;\;q^{\mu}_C &\sim (\lambda^{2},\lambda^{1},\lambda^{1},\lambda^{1})   \;,\nn  \\ 
  \text{for collinear sources:}\;\;\;q^{\mu}_G &\sim (\lambda^{2},\lambda^{1},\lambda^{1},\lambda^{2})  \;.
   \label{eq:glauber-momenta}
\end{align}

We now follow the discussion in Sec. 4.1 of Ref.~\cite{Ovanesyan:2011xy} and~\cite{Idilbi:2008vm} to establish the scaling of the gluon fields $A^{\mu}_{G}$ and $A^{\mu}_{C}$ for the three sources of the virtual gluons. Using Eqs. (4.2) and (4.3) along with the first row of Table 1 in Ref.~\cite{Ovanesyan:2011xy}, we establish the  scaling shown in Table~\ref{tb:scaling-A} of this paper. These scalings corresponds to the maximum allowed components for each source. For example Glauber scaling for soft and static sources is also kinematically allowed but the Lagrangian terms resulting from such scaling are power suppressed due to the phase-space integration for the sources.

\begin{table}[h!]
  \renewcommand{\arraystretch}{1.4}
  \begin{center}
    \begin{tabular}{|r|c|c|c|}
      \hline
      Source & Collinear      & Static       & Soft    \\ \hline \hline
      $\;\;A_C^{\mu} \sim $  &                 n.a.                       &$(\lambda^1,\lambda^2,\lambda^2,\lambda^2)$&$(\lambda^1,\lambda^1,\lambda^1,\lambda^1)$   \\ \hline
      $\;\;A_G^{\mu} \sim $  & $(\lambda^2,\lambda^3,\lambda^3,\lambda^2)$ &                 n.a.                     &               n.a.                           \\ \hline
    \end{tabular}
    \caption{The Glauber/Coulomb filed scaling for different sources of interaction in matter.}
    \label{tb:scaling-A}
  \end{center}
\end{table}

Since  we would often like to pick the dominant component for the momenta of the Glauber gluons, it  is useful to define
\begin{equation}
  \bmat{q}_{T} = (q_1,q_2,0)\; ,
\end{equation}
such that
\begin{equation}
  q^{\mu}_G = (0,\bmat{q}_T) + q^{\mu}_{us}\;,\;\;  \text{with}\;,\;\;  q_{us}^{\mu} \sim (\lambda^2,\lambda^2,\lambda^2,\lambda^2)\; ,
\end{equation}
and, similarly, for Coulomb gluons $q^{\mu}_C = (0,\bmat{q}) + q^{\mu}_{us}$.


\subsection{The background field method }
\label{sec:bfa}
We now proceed with the calculation of the Glauber/Coulomb and heavy quark interactions within the naive background filed method. Here, we shift the ultra-soft gluon fields in the vNRQCD Lagrangian in Eq.(\ref{eq:L-vNRQCD}): $A_{us}^{\mu,a} \to A_{us}^{\mu,a} + A_{G/C}^{\mu,a}$. After this shift, we read the interaction Lagrangian, $\mathcal{L}_{Q-G/C}$, from the leading expansion in $\lambda$ linear in $A_{G/C}^{\mu,a}$. As mentioned above, this approach is problematic and yields the wrong results. Nonetheless, we proceed with this exercise since it will help us set up the goals of the following section and, in addition, it demonstrates the dangers of not carefully consider the distinction of soft and ultra-soft scales.

We only consider the heavy quark sector, i.e. $\mathcal{L}_{Q-G/C}$, since the antiquark can follow trivially. We will organize the result by powers of $\lambda$,
\begin{equation}
  \mathcal{L}_{Q-G/C} =  \mathcal{L}_{Q-G/C}^{(0)} + \mathcal{L}_{Q-G/C}^{(1)} + \mathcal{L}_{Q-G/C}^{(2)} + \cdots \;,
\end{equation}
where if $\mathcal{L}_{Q-G/C}^{(0)}$ (for a particular source) scales as $\lambda^{m}$ then $\mathcal{L}_{Q-G/C}^{(n)} \sim \lambda^{m+n}$. For each source, in this paper, we will consider only the first two terms from the above equation, i.e. $ \mathcal{L}_{Q-G/C}^{(0)}$ and  $\mathcal{L}_{Q-G/C}^{(1)}$.  

Its clear from the form of the NRQCD Lagrangian and the scaling of the Glauber/Coulomb background fields  (Table~\ref{tb:scaling-A}) that the corrections to the leading Lagrangian from Glauber/Coulomb gluon exchanges have the following form,
\begin{equation}
  \label{eq:L0-BF}
  \mathcal{L}_{Q-G/C}^{(0)} (\psi,A_{G/C}^{\mu,a})  = \sum_{\bmat{p},\bmat{p}'}\psi^{\dag}_{\bmat{p}'} \lp  - g A_{G/C}^{0}(x) \rp \psi_{\bmat{p}}\;\; (collinear/static/soft).
\end{equation}
For the sub-leading Lagrangian we have contributions only from the collinear and soft sources:
\begin{align}
  \label{eq:L1-BF}
  \mathcal{L}_{Q-G}^{(1)} (\psi,A_{G}^{\mu,a}) & =  g \sum_{\bmat{p},\bmat{p}'} \psi^{\dag}_{\bmat{p}'} \lp \frac{ A^{\bmat{n}}_{G} \bmat{n}  \cdot \bmat{p}  }{m} \rp \psi_{\bmat{p}}\;\; (collinear), \nn\\
  \mathcal{L}_{Q-C}^{(1)} (\psi,A_{C}^{\mu,a})  &= 0\;\; (static) \;,  \nn\\
  \mathcal{L}_{Q-C}^{(1)} (\psi,A_{C}^{\mu,a}) & =  g \sum_{\bmat{p},\bmat{p}'} \psi^{\dag}_{\bmat{p}'} \lp \frac{  \bmat{A}_C  \cdot \bmat{p}  }{m} \rp \psi_{\bmat{p}}\;\; (soft),
\end{align}
where $A^{\bmat{n}} = \bmat{n} \cdot \bmat{A} $ and  $\bmat{n}$ is the collinear direction (in our convention $\bmat{n} = (0,0,1)$).
Note that,  for both $\mathcal{L}^{(0)}$ and $\mathcal{L}^{(1)}$,  the creation and annihilation of the heavy quark (or antiquark) are not evaluated at the same momenta, i.e. $\bmat{p} \neq \bmat{p}'$, since momentum is shifted by the Glauber/Coulomb gluon. This suggests that the naive shift of the fields might not yield the correct result due to the ambiguity in the choice of $\bmat{p}$ and $\bmat{p}'$ in the Lagrangian $\mathcal{L}^{(1)}$. Indeed, the correct $\mathcal{L}^{(1)}$ can be calculated in the non-relativistic limit of QCD with the hybrid and matching methods which we will discuss in the following section. In Appendix~\ref{app:last} we include a detailed discussion on how to properly implement the background field approach consistent with the power counting procedure. This, then will give results in agreement with the non-relativistic limit of QCD.


\section{Non-relativistic limit of QCD (NRQCD) }
\label{sec:F_rules}

To approach more systematically the inclusion of Glauber/Coulomb gluons in the NRQCD Lagrangian, we begin with some definitions and establishing the notation and conventions we will be using in the rest of this section. We then continue with an exercise to establish some of the terms of the known vNRQCD Lagrangian. This will help us to smoothly transition into the  main goal of this analysis, which is introducing the Glauber and Coulomb gluon interaction with the heavy quarks.


We will consider the leading and sub-leading  corrections to the NRQCD Lagrangian from Glauber and Coulomb gluon exchanges and start with fermonic sources (collinear, static, and soft). We will work in the chiral representation of Dirac matrices.
\begin{align}
  \gamma^{\mu} &=
  \begin{pmatrix}
    0 & \sigma^{\mu}\\
    \bar{\sigma}^{\mu} & 0
  \end{pmatrix}
  \,,&
  \text{where}\;\;\; \sigma^{\mu} &= (1,\bmat{\sigma})\;, \;\;\bar{\sigma}^{\mu} = (1,-\bmat{\sigma}) \;.
\end{align}
Then the Dirac spinors in this representation  take the following form:
\begin{align}
  u(p) &=
  \begin{pmatrix}
    \sqrt{p\cdot \sigma} \;\xi\\
    \sqrt{p\cdot \bar{\sigma}}\; \xi
  \end{pmatrix}
  \,,&
  v(p) &=
  \begin{pmatrix}
    \sqrt{p\cdot \sigma} \;\eta \\
    -\sqrt{p\cdot \bar{\sigma}}\; \eta
  \end{pmatrix} \;.
\end{align}
The non-relativistic limit of those ($\vert \bmat{p} \vert \ll p_0$) is  given by
\begin{align}
  \label{eq:spinors-BST}
  u(p) &= \sqrt{p_0}\lp 1 - \frac{\bmat{p} \cdot \bmat{\gamma}}{2 p_0} - \frac{\bmat{p}^2}{8 p_0^2} +\cdots \rp u^{(0)}\;, &  v(p)= \sqrt{p_0}\lp 1 + \frac{\bmat{p} \cdot \bmat{\gamma}}{2 p_0} - \frac{\bmat{p}^2}{8p_0^2} + \cdots \rp v^{(0)} \;,
\end{align}
where the ellipsis denotes terms of higher order in $\vert \bmat{p} \vert / p_0$. The normalized rest frame spinors $u^{(0)}$ and $v^{(0)}$ are given by
\begin{align}
  \label{eq:spinors-RF}
  u^{(0)} &= 
  \begin{pmatrix}
    \xi\\
    \xi
  \end{pmatrix}
  \,,&
  v^{(0)} &= 
  \begin{pmatrix}
    \eta \\
    -\eta
  \end{pmatrix} \;,
\end{align}
and satisfy the equations of motion
\begin{align}
  \label{eq:EoM-RF}
  (1-\slashed{v}) u^{(0)} &= 0\;,& (1+\slashed{v}) v^{(0)} &= 0 \;,
\end{align}
with $v^{\mu} = (1, \bmat{0})$. 

\subsection{Interactions with ultra-soft gluons}
In this subsection we will show how one can reconstruct the tree-level NRQCD Lagrangian involving single ultra soft gluon interactions with the heavy quarks.  In this exercise we will build the formalism and all ingredients necessary  to introduce the Glauber and Coulomb gluon interactions. We do that by studying the non-relativistic limit of the expectation value of the QCD operator $\mathcal{O}_{1}$,
\begin{equation}
  \mathcal{O}_1 = \int d^4x \;\bar{\Psi} \lp i \slashed{\partial} -g \slashed{A} - m   \rp \Psi (x) \;.
\end{equation}
 We will consider the single particle expectation value of the operator $\mathcal{O}_1$ for extracting the kinematic terms in the NRQCD Lagrangian,
\begin{equation}
  \label{eq:kinematic}
  \lp \includegraphics[width=0.15\linewidth, valign=c]{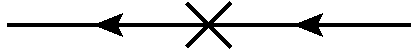} \rp
  _{\text{QCD}(\lambda \ll 1 )} = \;\;\;  \includegraphics[width=0.15\linewidth, valign=c]{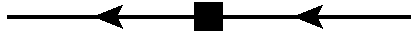} \;,
\end{equation}
where we interpret the RHS of the above diagrammatic equation as the corresponding terms generated by the non-relativistic version of $\mathcal{O}_1$. Similarly, for  the interaction terms we then consider an expectation value where the initial state contains an additional gluon. This corresponds to,
\begin{equation}
  \label{eq:interactions}
  \lp \includegraphics[width=0.15\linewidth, valign=c]{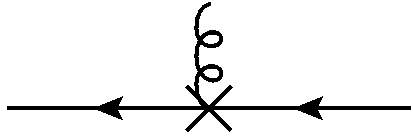} \rp_{\text{QCD}(\lambda \ll 1 )} = \;\;\;  \includegraphics[width=0.15\linewidth, valign=c]{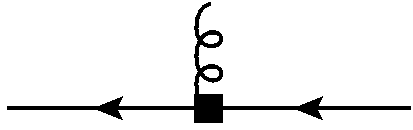} \;.
\end{equation}
In principle, in the above equation we need to consider insertions from the QCD Lagrangian in the LHS and the corresponding NRQCD contributions in the RHS. Its easy to demonstrate that, including those terms and after some simplifications, the result reduced to the same equation as above.

We start with the kinematic terms in Eq.~(\ref{eq:kinematic}).
\begin{equation}
  \label{eq:EoM}
  \includegraphics[width=0.15\linewidth, valign=c]{O1}\;  = \Langle Q(p') \Bvert \mathcal{O}_1 \Bvert Q(p) \Rangle = \underbrace{\bar{u}(p')\lp \slashed{p} - m \rp u(p)}_{{ }_{ \textstyle{\equiv V_{2Q} (p,p')} }} \; \delta^{(4)}(r - r')\;.
\end{equation}
The RHS of  Eq.~(\ref{eq:EoM}) vanishes from the equation of motion (EoM), but instead of applying EoM, we will first take the non-relativistic limit which will give the corresponding EoM for the non-relativistic heavy quark (i.e. Schr\"{o}dinger's equation for free particles). To better understand this statement, imagine a function $f(\lambda)$ that depends on a small parameter $\lambda$. If the function vanishes for all values of $1 > \lambda > 0$, then if we expand in powers of $\lambda$ the coefficients have to vanish independently. In the context of NRQCD, $\lambda$ is the velocity of the heavy quark and we are interested in the leading non-trivial coefficient. Since all coefficients vanish,  by non-trivial we mean that an additional condition needs to be imposed for them to vanish. We then interpret this condition as the equation of motion for the non-relativistic theory. Alternatively, one may add a small offshellness to the momenta $p$ and $p'$ using $r_0 \to \tilde{r}_0$ and $r'_0 \to \tilde{r}'_0$. Then the first non-vanishing term is what we are after. 

In Eq.~(\ref{eq:interactions}) we have not yet specified the scaling of the vector field or its momenta. For constructing the vNRQCD Lagrangian we will take this gluon to be ultra-soft,
\begin{equation}
  \label{eq:QAQ}
  \includegraphics[width=0.15\linewidth, valign=c]{O1-A}\;= \Langle Q(p') \Bvert \mathcal{O}_1 \Bvert Q(p) + g(q)\Rangle =  \underbrace{- \bar{u}(p')\lp g  \slashed{A}_U (q) \rp u(p)}_{{ }_{ \textstyle{\equiv V_{2Q,A}(p,q,p')} }} \; \delta^{(4)}(r + q - r') \;,
\end{equation}
where $g(q)$ is an ultra-soft gluon with momenta $q \sim (\lambda^2,\lambda^2,\lambda^2,\lambda^2)$. We take the non-relativistic limit of Eq.~(\ref{eq:EoM}) by expanding up-to the leading correction the spinors,  and up-to the subleading propagator. For this, we use the Eqs.~(\ref{eq:spinors-BST}), (\ref{eq:spinors-RF}), and (\ref{eq:momenta-deco}). We explicitly show all steps.\\\\
\textbullet $\;\;\mathcal{O}(\lambda^0)$: At leading power (LP), we expand all relevant elements only in the leading velocity terms, that is the absolute non-relativistic limit where the heavy quark is at rest:
  \begin{equation}
    V_{2Q}^{(0)} = -m(u^{(0)})^{\dag} \gamma^0 (1-\slashed{v}) u^{(0)} = 0 \;,
  \end{equation}
  which vanished using Eq.~(\ref{eq:EoM-RF}).\\
\textbullet $\;\;\mathcal{O}(\lambda^1)$: The next-to-leading power (NLP) expansion we represent using the residual components $r$ and $r'$ as defined in  Eqs.(\ref{eq:momenta-deco}), (\ref{eq:deco2}), and (\ref{eq:deco1-deco2}):
  \begin{equation}
    V_{2Q}^{(1)} = -m(u^{(0)})^{\dag} \lbc \lp\frac{\bmat{r}' \cdot \bmat{\gamma}}{2m} \rp \gamma^0 (1-\slashed{v}) +   \gamma^0 \lp\frac{\bmat{r}' \cdot \bmat{\gamma}}{m} \rp -  \gamma^0 (1-\slashed{v}) \lp\frac{\bmat{r} \cdot \bmat{\gamma}}{2m} \rp  \rbc u^{(0)} = 0 \;.
  \end{equation}
  Each of the three terms in curly brackets comes from expanding at leading order one of the following: $\bar{u}(p'), (\slashed{p} -m),$ and $u(p)$. All three terms vanish independently. We will see later that this is a consequence of what we will define as the \emph{equation of odd gammas}. \\   
\textbullet $\;\;\mathcal{O}(\lambda^2)$: For the next-to-next-to-leading power (NNLP) expansion we need the $\mathcal{O}(\bmat{r}^2/m^2)$ from each of $\bar{u}(p'), (\slashed{p} -m),$ and $u(p)$ but also contributions from mixed NLP expansion:
  \begin{multline}
    V_{2Q}^{(2)} = -m(u^{(0)})^{\dag} \lbc  \lp  \frac{r'_0}{2m} -\frac{\bmat{r}'^2 }{8m^2}  \rp \gamma^0 (1-\slashed{v}) - \frac{r_0}{m} + \lp \frac{r_0}{2m} -\frac{\bmat{r}^2 }{8m^2}  \rp \gamma^0 (1-\slashed{v})   \rbc u^{(0)}  \\
    -m(u^{(0)})^{\dag} \lbc \lp\frac{\bmat{r}' \cdot \bmat{\gamma}}{2m} \rp \gamma^0 \lp\frac{\bmat{r} \cdot \bmat{\gamma}}{m} \rp -   \gamma^0  \lp\frac{\bmat{r} \cdot \bmat{\gamma}}{m} \rp   \lp\frac{\bmat{r} \cdot \bmat{\gamma}}{2m} \rp -    \lp\frac{\bmat{r}' \cdot \bmat{\gamma}}{2m} \rp \gamma^0  (1-\slashed{v}) \lp\frac{\bmat{r} \cdot \bmat{\gamma}}{2m} \rp  \rbc u^{(0)} \;.
  \end{multline}
  To simplify this result we note that the first and last term in the curly brackets of the first line, vanish from application of Eq.~(\ref{eq:EoM-RF}). To simplify the second line we use:
  \begin{align}
    (1-\slashed{v}) \bmat{\gamma}  &= \bmat{\gamma} (1+\slashed{v})\;, & (1+\slashed{v})u^{(0)} &= 2u^{(0)}\;, & (u^{(0)})^{\dag} \gamma^0 =  (u^{(0)})^{\dag} \;.
  \end{align}
  With these modifications the result significantly simplifies to give a familiar expression,
  \begin{equation}
    \label{eq:V2-final}
    V_{2Q}^{(2)} = ( \sqrt{2m} \xi^{\dag} ) \lbc  r_0 - \frac{\bmat{r}^2}{2m} \rbc( \sqrt{2m} \xi ) \;.
  \end{equation}
  Since $V_{2Q}^{(2)}$ need to vanish, then $r_0 = \bmat{r}^2/2m$, which is exactly the well-known non-relativistic relation between the kinetic energy and the  three-momenta.\\
\textbullet $\;\;\mathcal{O}(\lambda^3)$: All terms that contribute to this order can  easily be shown to have one or three $\gamma^i$ squeezed between the  $(u^{(0)})^{\dag}$ and $u^{(0)}$. This means that all of them vanish.  This statement can be generalized to any odd power, $n$, of $\gamma^i$: 
  \begin{align}
    (u^{(0)})^{\dag} \gamma^{i_1} \gamma^{i_2} \cdots \gamma^{i_n} u^{(0)} = -(-1)^{\frac{n+1}{2}} (u^{(0)})^{\dag}
    \begin{pmatrix}
      0 & \sigma^{i_1} \sigma^{i_2} \cdots \sigma^{i_n}\\
      - \sigma^{i_1} \sigma^{i_2} \cdots \sigma^{i_n} & 0
    \end{pmatrix} u^{(0)}  = 0 \; .
  \end{align}
  For future reference we will refer to the above equation as the {\it equation of odd gammas}. Thus:
  \begin{equation}
    V_{2Q}^{(3)} = 0  \;.
  \end{equation}
In order to account for the $\mathcal{O}(\lambda^3)$ terms that come for the decomposition of soft and ultra-soft  (see in Eq.~(\ref{eq:deco2})), we need to make replacements as described in Eq.~(\ref{eq:deco1-deco2}). This will give for the leading and subleading contributions,
\begin{equation}
  \includegraphics[width=0.15\linewidth, valign=c]{O1-NR} \; = ( \sqrt{2m} \xi^{\dag} ) \lbc  r_{0,us} - \frac{(\bmat{r}_s + \bmat{r}_{us})^2}{2m} \rbc( \sqrt{2m} \xi )\; \delta^{(4)} (r_{us} - r_{us}')\;\delta_{\bmat{r},\bmat{r}'} \;.
\end{equation}
We can now write the Lagrangian that would generate such term,
\begin{equation}
  \mathcal{L}^{free}_{\text{NRQCD}} = \sum_{\bmat p} \psi_{\bmat{p}}^{\dag} \lp i \partial_t -\frac{(\bmat{\mathcal{P}} - i \bmat{\partial})^2 }{2m} \rp \psi_{\bmat p} + \mathcal{O}(\lambda^4) \;.
\end{equation}
We kept the term proportional to $\bmat{\partial}^{2}$ even though is of higher order ($\mathcal{O}(\lambda^4)$) than what we are considering here. This will later help us write the final Lagrangian in a gauge invariant form. In the above equation, $\psi_{\bmat{p}}(x)$ is the two-component Pauli spinor that  satisfy the two-component Schr\"{o}dinger's equation:
\begin{equation}
  \lp i \partial_t -\frac{\bmat{\mathcal{P}}^2}{2m} \rp \psi_{\bmat p} = 0\;.
\end{equation}

We now turn to the $V_{2Q,A}$. Since $A_{U}^\mu$ is an ultra-soft gluon we have,
\begin{equation}
  A_{U}^\mu \sim (\lambda^2,\lambda^2,\lambda^2,\lambda^2) \;,
\end{equation}
and thus our expansion of $V_{2Q,A}$ starts from $\mathcal{O}(\lambda^2)$, compared to $V_{2Q}^{(0)}$.\\\\
\textbullet $\;\;\mathcal{O}(\lambda^2)$:  This result, we can trivially get from the LP expansion of $\bar{u}(p)$ and $u(p)$.
  \begin{equation}
    \label{eq:V2.1-2QA}
    V^{(2)}_{2Q,A} = - m g (u^{(0)})^{\dag }\lp \gamma^0   \slashed{A}_U  \rp u^{(0)} \;.
  \end{equation}
  Then from the equation of odd gammas we have
  \begin{equation}
    \label{eq:V2.2-2QA}
    V^{(2)}_{2Q,A} = - m g (u^{(0)})^{\dag }\lp  \gamma^{0} A^{0}_U  \rp u^{(0)} = -(\sqrt{2m} \xi^{\dag}) \lp g A^{0}_{U} \rp (\sqrt{2m} \xi) \;.
  \end{equation}
\textbullet $\;\;\mathcal{O}(\lambda^3)$:  We would like to utilize the result we get in this section later, when we extent to Glauber and Coulomb regions instead of ultra-soft. For this reason, we work with generic three-momenta   and we will implement the momentum conservation delta function at the end,
  \begin{equation}
    V^{(3)}_{2Q,A} = - m g (u^{(0)})^{\dag }\lbc \lp\frac{\bmat{r}' \cdot \bmat{\gamma}}{2m} \rp  \slashed{A}_U     - \gamma^0  \slashed{A}_U    \lp\frac{\bmat{r} \cdot \bmat{\gamma}}{2m} \rp  \rbc u^{(0)} \;.
  \end{equation}
  Again, from the equation of odd gammas only the $\mu =k= \{1,2,3\}$ will contribute to this result
  \begin{align}
    \label{eq:V3-2QA-total}
    V^{(3)}_{2Q,A} &= - m g (u^{(0)})^{\dag }\lbc  \lp\frac{\bmat{r}' \cdot \bmat{\gamma}}{2m} \rp   \bmat{\gamma} \cdot \bmat{A}_U     + \bmat{\gamma} \cdot \bmat{A}_U  \lp\frac{\bmat{r} \cdot \bmat{\gamma}}{2m} \rp  \rbc u^{(0)} \nn\\
    &=- m g (u^{(0)})^{\dag }\lbc \gamma^i \gamma^k   \rbc u^{(0)} \; \lp \frac{(r')^i A^k_{U} + r^k A^i_{U}}{2m} \rp \nn\\
    &=+m g (\sqrt{2m} \xi^{\dag}) \lbc \sigma^{i}\sigma^{k} \rbc (\sqrt{2m} \xi)\; \lp \frac{(r')^i A^k_{U} + r^k A^i_{U}}{2m} \rp \nn\\
    &=\frac{g}{2m} (\sqrt{2m} \xi^{\dag}) \lbc \bmat{A}_U \cdot (\bmat{r}'+\bmat{r}) - i \lp \bmat{A}_U \times(\bmat{r'} - \bmat{r} )\rp \cdot \bmat{\sigma}  \rbc (\sqrt{2m} \xi) \;.
  \end{align}
  Using the  momentum conservation delta function and expanding $\bmat{r}$ in its soft and ultra-soft components we get 
  \begin{equation}
    \label{eq:V3-2QA}
    V^{(3)}_{2Q,A} = \frac{g}{2m} (\sqrt{2m} \xi^{\dag}) \lbc \bmat{A}_U \cdot (2\bmat{r}_s+2\bmat{r}_{us}+\bmat{q}) - i \lp \bmat{A}_U  \times \bmat{q} \rp \cdot \bmat{\sigma}  \rbc (\sqrt{2m} \xi)\;.
  \end{equation}

We now have all the ingredients to construct the interaction Lagrangian of NRQCD up-to corrections of $\mathcal{O}(\lambda^{3})$. Adding the two terms together
\begin{equation}
  \includegraphics[width=0.15\linewidth, valign=c]{O1-A-NR} \; = g (\sqrt{2m} \xi^{\dag}) \lbc - A^{0}_{U}+ \frac{\bmat{A}_U \cdot \bmat (2\bmat{r}_s+2\bmat{r}_{us}+\bmat{q})}{2m} \rbc (\sqrt{2m} \xi)\; \delta^{(4)}(r_{us}+q-r_{us}')\;\delta_{\bmat{r},\bmat{r}'} \;.
\end{equation}
The term $2\bmat{r}_{us} + \bmat{q}$ is of $\mathcal{O}(\lambda^4)$ but we keep it anyway because will help to write the Lagrangian in a gauge invariant form. We, thus, have
\begin{equation}
  \mathcal{L}_{\text{NRQCD}}^{int.} = \sum_\bmat{p} \psi_\bmat{p}^{\dag} \lp  - g A^{0}_{U}+ \frac{ 2\bmat{A}_U\cdot (\bmat{\mathcal{P}} - i \bmat{\partial}) -i (\bmat{\partial} \cdot \bmat{A}_U)  }{2m} \rp \psi_{\bmat{p}} +\mathcal{O}(\lambda^4) \;.
\end{equation}
Therefore, for the total Lagrangian we obtain
\begin{equation}
  \mathcal{L}_{\text{NRQCD}} =\mathcal{L}_{\text{NRQCD}}^{free}+\mathcal{L}_{\text{NRQCD}}^{int.} = \sum_\bmat{p} \psi_\bmat{p}^{\dag} \lp  iD^{0}_{U} - \frac{ (\bmat{\mathcal{P}} -i \bmat{D}_{U} )^2}{2m} \rp \psi_{\bmat{p}} +\mathcal{O}(\lambda^4) \;,
\end{equation}
where we have introduced an $\mathcal{O}(\lambda^4)$ term, quadratic in the vector field $\bmat{A}$, such that we can write the Lagrangian in a gauge invariant form. The interaction terms we constructed here involve only a single gluon vertex. Larger number of gluons contribute only at $\mathcal{O}(\lambda^{4})$ and higher. For example, from conservation of momentum the difference of the three momentum of the in and out heavy quark is simply the ultra-soft component of the gluon. Of course,  up-to the order we are working here this contribution is not relevant, but if we have kept this term we would have, 
\begin{equation}
  \bmat{A}_U \times(\bmat{r'} - \bmat{r} ) = \bmat{A}_U \times \bmat{q} \;.
\end{equation}
This corresponds to a term in the Lagrangian of the form
\begin{equation}
  \sum_{\bmat{p}} \frac{g}{2m}\psi_{\bmat{p}}^{\dag} \lp i\bmat{\partial}\times \bmat{A}  \rp \psi_{\bmat{p}} \;,
\end{equation}
which is the abelian part of the chomomagnetic operator $B^i =  \epsilon^{ijk} G^{jk}/2$. The complete chromo-magnetic operator contains  also a non-abelian part with two gluon fields which we do not reproduce here, but they can be introduced through gauge completion. Alternatively, one can explicitly calculate the contribution of the terms quadratic in the vector field by evaluating the following:
\begin{equation}
  \label{eq:2A-diagrams}
  \includegraphics[width=0.15\linewidth, valign=c]{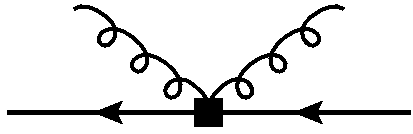} \; = \; \lp \includegraphics[width=0.15\linewidth, valign=c]{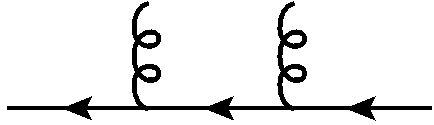} + \text{perm.} \rp_{\text{QCD}(\lambda \ll 1)} - \lp \includegraphics[width=0.15\linewidth, valign=c]{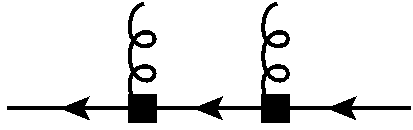} + \text{perm.} \rp \;,
\end{equation}
where is understood that in the RHS the first term is to be evaluated in the non-relativistic limit. The subtraction of the NRQCD diagram is necessary to avoid double counting. We will no further pursue this analysis here.

\subsection{Introducing the Glauber and Coulomb interactions }

Here we introduce the Glauber/Coulomb interactions by repeating the analysis of expanding in $\lambda$ the $\mathcal{O}_1$ expectation value $V_{2Q,A}$, but this time assuming Glauber/Coulomb gluon scaling instead of ultra-soft. This approach we refer to as hybrid method. The relevant scalings that control the power-counting expansion are then given by Eq.~(\ref{eq:glauber-momenta}) and Table~\ref{tb:scaling-A}. To simplify the discussion we will utilize many of the results from the last subsection. \\\\
\textbullet $\;\;\mathcal{L}^{(0)}$: We can use the results from Eqs.~(\ref{eq:V2.1-2QA}) and (\ref{eq:V2.2-2QA}) and directly get:
  \begin{equation}
    \label{eq:V2.2-2QAG}
    V^{(2)}_{2Q,A_{G/C}} = -(\sqrt{2m} \xi^{\dag}) \lp g A^{0}_{G/C} \rp (\sqrt{2m} \xi) \;.
  \end{equation}
\textbullet $\;\;\mathcal{L}^{(1)}$:  We utilize the final expression for $V^{(3)}_{2Q,A}$ from the last line of Eq.~(\ref{eq:V3-2QA}) and, performing the proper power-counting for $\bmat{q}$, we have:
  \begin{equation}
    V^{(3)}_{2Q,A_{G/C}}=\frac{g}{2m} (\sqrt{2m} \xi^{\dag}) \lbc \bmat{A}_{G/C} \cdot (2\bmat{r}_s+\bmat{q}) + i \lp \bmat{q} \times \bmat{A}_{G/C} \rp \cdot \bmat{\sigma}  \rbc (\sqrt{2m} \xi) \;.
  \end{equation}
  Since the components $A^{i}_{G/C}$ for $i=1,2,3$ have different scaling for each source, in order to continue we need to specify the source of the Glauber/Coulomb gluon.\\
  \underline{{\it  Collinear:}} \\
  \begin{equation}
    V^{(3),coll.}_{2Q,A_{G}}=\frac{g}{2m} (\sqrt{2m} \xi^{\dag}) A^{\bmat{n}}_{G }\lbc 2\; \bmat{n} \cdot \bmat{r}_s + i \lp \bmat{q}_T \times \bmat{n} \rp \cdot \bmat{\sigma}  \rbc (\sqrt{2m} \xi)\;,
  \end{equation}
  \underline{{\it  Static:}} \\
  \begin{equation}
    V^{(3),stat.}_{2Q,A_{C}} = 0 \;,
  \end{equation}
  \underline{{\it  Soft:}} \\
  \begin{equation}
    V^{(3),soft}_{2Q,A_{C}}=\frac{g}{2m} (\sqrt{2m} \xi^{\dag}) \lbc \bmat{A}_{C} \cdot (2\bmat{r}_s+\bmat{q}) + i \lp \bmat{q} \times \bmat{A}_{C} \rp \cdot \bmat{\sigma}  \rbc (\sqrt{2m} \xi) \;.
  \end{equation}

We are now ready to write the leading and subleading correction to the NRQCD$_{\rm G}$ Lagrangian in the heavy quark sector from virtual (Glauber/Coulomb) gluon insertions, i.e. $\mathcal{L}_{Q-G}$, :
\begin{equation}
  \label{eq:L0-NR}
  \mathcal{L}_{Q-G/C}^{(0)} (\psi,A_{G/C}^{\mu,a})  = \sum_{\bmat{p},\bmat{q}_T}\psi^{\dag}_{\bmat{p}+\bmat{q}_T} \lp - g A^{0}_{G/C} \rp \psi_{\bmat{p}}\;\; (collinear/static/soft)\; , 
\end{equation}
and 
\begin{align}
  \label{eq:L1-NR}
  \mathcal{L}_{Q-G}^{(1)} (\psi,A_{G}^{\mu,a}) & =  g\sum_{\bmat{p},\bmat{q}_T} \psi^{\dag}_{\bmat{p} +\bmat{q}_T} \lp\frac{ 2  A_{G}^\bmat{n} (\bmat{n} \cdot \bmat{\mathcal{P}}) - i \lb ( \bmat{\mathcal{P}}_{\perp} \times \bmat{n}) A^{\bmat{n}}_{G}  \rb \cdot \bmat \sigma }{2m} \rp \psi_{\bmat{p}}\;\; (collinear)\; , \nn\\
  \mathcal{L}_{Q-C}^{(1)} (\psi,A_{C}^{\mu,a})  &= 0\;\; (static)\;,\nn\\
  \mathcal{L}_{Q-C}^{(1)} (\psi,A_{C}^{\mu,a}) & =  g\sum_{\bmat{p},\bmat{q}_T} \psi^{\dag}_{\bmat{p} +\bmat{q}_T} \lp\frac{ 2  \bmat{A}_{C} \cdot \bmat{\mathcal{P}} + [\bmat{\mathcal{P}} \cdot  \bmat{A}_{C} ] - i \lb  \bmat{\mathcal{P}} \times \bmat{A}_{C} \rb \cdot \bmat \sigma }{2m} \rp \psi_{\bmat{p}}\;\; (soft)\;,
\end{align}
where we use squared brackets in order to denote the region in which the label momentum operator, $\mathcal{P}^{\mu}$, acts. Eqs.~(\ref{eq:L0-NR}) and (\ref{eq:L1-NR}) are the main results of this section. Comparing to the corresponding result from the background field approach in Eqs.~(\ref{eq:L0-BF}) and (\ref{eq:L1-BF}), we see that the results for the  leading Lagrangian, $\mathcal{L}_{Q-G/C}^{(0)}$ agree. For the subleading Lagrangian, $\mathcal{L}_{Q-G/C}^{(1)}$, we find that for the cases of collinear and soft sources there are additional terms that appeared in the hybrid method. We further discuss the origin of the discrepancy in Appendix~\ref{app:last}.


\subsection{Matching from QCD including source fields  }
Here, we will reproduce the results in  Eqs.~(\ref{eq:L0-NR}) and (\ref{eq:L1-NR}) by considering the non-relativistic limit of the $t$-channel diagram for a particular source. We consider both quark and gluon sources. This will give the fields $A_G$ and $A_C$, appearing in  Eqs.~(\ref{eq:L0-NR}) and (\ref{eq:L1-NR}), as a function of the source currents. We begin with the collinear quark source
\begin{multline}
  t_{q-coll.}=\;\tensor*[^{\textstyle{p'}\;}_{\textstyle{p_n'}\;}]{\includegraphics[width=0.15\linewidth, valign=c]{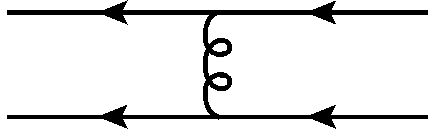} }{^{\;\textstyle{p}}_{\;\textstyle{p_n}}}\;= i \bar{u}(p')(g\gamma^{\mu} T^a) u(p)\frac{ g_{\mu\nu}}{q^2} \bar{u}(p_n')(g\gamma^{\nu} T^a) u(p_n) \\ = t_{q-coll.}^{(0)} + t_{q-coll.}^{(1)} + \mathcal{O}(\lambda^2)\;,
\end{multline}
where  $p_n$ and $p_n'$ are the momenta of the incoming and outgoing collinear quarks, respectively,  and $p$ and $p'$ are the momenta of the corresponding heavy quarks. Taking the collinear limit for the spinor $u(p_n)$ and the non-relativistic limit for $u(p)$ we get
\begin{equation}
  \label{eq:t0-q-coll}
  t^{(0)}_{coll.}=(\sqrt{2m}\xi^{\dag}) (-ig v^{\mu} T^a) (\sqrt{2m}\xi) \lp \frac{n_{\mu}}{\bmat{q}_T^2} \bar{u}_{n}(p_n) (g T^a) \frac{\slashed{\bar{n}}}{2} u_n(p_n) \rp \;.
\end{equation}
We then interpret this term as a Feynman diagram generated by the following Lagrangian:
\begin{align}
  \label{eq:L0-t-q}
  \mathcal{L}_{Q-G}^{(0)} (\psi,A^{\mu,a}_G) & = \sum_{\bmat{p},\bmat{q}_T}  \psi^{\dag}_{\bmat{p}+\bmat{q}_T} \lp  - g T^a v_{\mu} \rp \psi_{\bmat{p}}\; A^{\mu,a}_G   \;, &
  \text{where}\;\;\;  A^{\mu,a}_G & =  \frac{n^{\mu}}{\bmat{q}_{T}^2} \sum_{\ell} \bar{\xi}_{n,\ell-\bmat{q}_T} \frac{ \slashed{\bar{n}}}{2} (g T^a)  \xi_{n,\ell} \;.
\end{align}
In the above equation  $n^{\mu} = (1,0,0,1)$ and $\bar{n}^{\mu} = (1,0,0,-1)$. This is exactly the result we obtained in Eq.~(\ref{eq:L0-NR}), but now we have an expression for the background Glauber gluon as a function of the source fields. For the next order result, $t^{(1)}_{coll.}$, we will keep the expansion of the collinear sector up-to the leading accuracy end expand the heavy quark spinors one order higher in the non-relativistic limit. For that we can utilize the result of Eq.~(\ref{eq:V3-2QA-total}) to write:
\begin{equation}
  \label{eq:heavy-expansion-1}
  - i m g (u^{(0)})^{\dag }\lbc  \lp\frac{\bmat{r}' \cdot \bmat{\gamma}}{2m} \rp   \bmat{\gamma}     + \bmat{\gamma}  \lp\frac{\bmat{r} \cdot \bmat{\gamma}}{2m} \rp  \rbc u^{(0)} 
  =\frac{ig}{2m} (\sqrt{2m} \xi^{\dag}) \lbc (\bmat{r}'+\bmat{r}) + i (\bmat{r'} - \bmat{r} ) \times \bmat{\sigma}  \rbc (\sqrt{2m} \xi)\;,
\end{equation}
then we have
\begin{equation}
  \label{eq:t1-q-coll}
  t^{(1)}_{q-coll.}= \lp \frac{ig}{2m} (\sqrt{2m} \xi^{\dag}) \lbc (2 \bmat{r}_s+\bmat{q}_T) - i \bmat{q}_T \times \bmat{\sigma}  \rbc T^{a} (\sqrt{2m} \xi)\rp \cdot \lp \frac{\bmat{n}}{\bmat{q}_T^2} \bar{u}_{n}(p_n) (g T^a) \frac{\slashed{\bar{n}}}{2} u_n(p_n) \rp \;.
\end{equation}
This is the result we get using he Lagrangian terms $\mathcal{L}_{Q-G}^{(1)}$ given in Eq.(\ref{eq:L1-NR}), with $A_{G}^{\mu,a}$ given by Eq.(\ref{eq:L0-t-q}). Since the non-relativistic expansion of the heavy spinors is independent of the sources, it is easy to extent this result for soft and static sources by simply performing the  following replacements:
\begin{align}
  Static:& \;\;\; \frac{-i g_{\mu\nu}}{q^2} \bar{u}(p_s')(-ig\gamma^{\nu} T^a) u(p_s) \;\to\; \frac{ v^{\mu}}{\bmat{q}^2}  (\sqrt{2m} \xi^{\dag})  (g T^a)  (\sqrt{2m} \xi^{\dag})\;,  \nn \\ 
  Soft:&\;\;\;    \frac{-i g_{\mu\nu}}{q^2} \bar{u}(p_s')(-ig\gamma^{\nu} T^a) u(p_s) \;\to\; \frac{1}{\bmat{q} ^2}    \bar{u}(p_s')  \gamma^{\mu} (g T^a)  u(p_s)\;.
\end{align}
With these substitutions,  and using the expansion in Eq.~(\ref{eq:heavy-expansion-1}), we find for the $t$-channel diagram with soft fermion source:
\begin{align}
  \label{eq:soft-t}
  t^{(0)}_{q-soft}=& (\sqrt{2m}\xi^{\dag}) (-ig v_{\mu} T^a) (\sqrt{2m}\xi) \, \lp \frac{1}{\bmat{q} ^2}  \bar{u}(p_s')  \gamma^{\mu} (g T^a)  u(p_s)   \rp   \;,  \nn \\ 
  t^{(1)}_{q-soft}=& \lp \frac{ig}{2m} (\sqrt{2m} \xi^{\dag}) \lbc (2 \bmat{r}_s+\bmat{q}) - i \bmat{q} \times \bmat{\sigma}  \rbc T^a (\sqrt{2m} \xi)\rp \cdot \lp   \frac{1}{\bmat{q} ^2}    \bar{u}(p_s') \bmat{\gamma} (g T^a)  u(p_s)  \rp  \;,
\end{align}
and with static fermion source,
\begin{align}
  t^{(0)}_{q-stat.}= &  (\sqrt{2m}\xi^{\dag}) (-ig v_{\mu} T^a) (\sqrt{2m}\xi) \, \lp  \frac{ v^{\mu}}{\bmat{q}^2}  (\sqrt{2m} \xi^{\dag})  (g T^a)  (\sqrt{2m} \xi^{\dag})  \rp \;,  \nn \\ 
  t^{(1)}_{q-stat.}=&\; 0 \;.
\end{align}
Is easy now to see how these terms for $t^{(0)}$ and $t^{(1)}$ are reproducing exactly the Lagrangian terms in Eqs.~(\ref{eq:L0-NR}) and  (\ref{eq:L1-NR}) with
\begin{equation}
  A_{C}^{\mu,a} \equiv \frac{ v^{\mu}}{\bmat{q}^2} \sum_{\bmat{\ell}}           \bar{h}_{v,\bmat{\ell}-\bmat{q}}  (g T^a)  h_{v,\bmat{\ell}}  \;,
\end{equation}
for a static source and 
\begin{equation}
  A_{C}^{\mu,a} \equiv \frac{1}{\bmat{q} ^2}     \sum_{\ell}                    \bar{\phi}_{\ell-\bmat{q}}  \gamma^{\mu} (g T^A)  \phi_{\ell} \;,
\end{equation}
for a soft source, where soft fermion fields $\phi_{\ell}$ are the same that appear in the vNRQCD Lagrangian in Eq.(\ref{eq:L-vNRQCD}), and $h_{v,\ell}$ are the heavy fermion field and its properties are governed by the HQET Lagrangian~\cite{Isgur:1989vq, Isgur:1989ed}.

Next, we consider gluon field sources. In this case, in addition to the $t$-channel diagram we have additional two diagrams that contribute to the same process. These two diagrams correspond to absorbing and re-emitting a collinear (or soft) gluon and are necessary to establish a full gauge invariant result when considering all polarizations of the propagating gluons. As before, we begin with the analysis of collinear sources,
\begin{eqnarray}
  \label{eq:g-diagrams}
  t_{g-coll.}&=&\;\tensor*[^{\textstyle{p'}\;}_{\textstyle{p_n'}}]{\includegraphics[width=0.15\linewidth, valign=c]{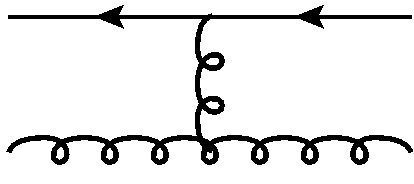} }{^{\;\textstyle{p}}_{\;\textstyle{p_n}}}\; + {\includegraphics[width=0.15\linewidth, valign=c]{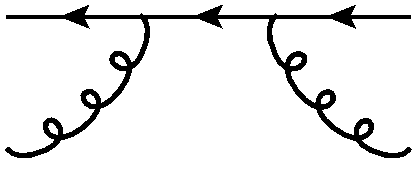} } + {\includegraphics[width=0.15\linewidth, valign=c]{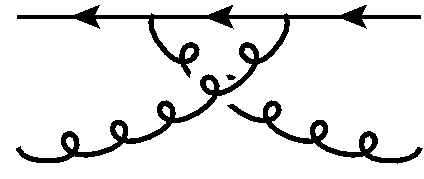} }  \nonumber  \\[2ex]
  &= & t^{(0)}_{g-coll.} + t^{(1)}_{g-coll.} + \mathcal{O}(\lambda^2) \; . 
\end{eqnarray}
Using the following power counting for the light-cone components (along the $n^{\mu}$ direction) of the collinear fields,
\begin{equation}
  A^{a,\mu}_n = (A^{+,a}_n,A^{-,a}_n,\bmat{A}^{a}_{n \perp}) \sim (\lambda^2,1,\lambda) \; , 
  \end{equation}
we expanding the spinors and the heavy quark propagators in the power-counting parameter $\lambda$ to get for the leading contribution:\begin{equation}
  \label{eq:t0-g-coll}
  t^{(0)}_{g-coll.} = g^2 f^{abc}(2m \xi^{\dag} T^{c} \xi) \; \lb  \frac{p_n^{-}}{\bmat{q}_T^{2}}\; \bmat{B}^{a (0)}_{n\perp, p_n} \cdot \bmat{B}^{b (0)}_{n \perp, p_n'} \rb  \;, 
\end{equation}
where 
\begin{equation}
  \bmat{B}^{a,(0)}_{n\perp,\ell}  \equiv \bmat{A}^{a}_{n\perp,\ell}  -  \bmat{p}_{n\perp} \frac{ A^{-,a}_{n,\ell} }{p_{n}^-} \; .
\end{equation}
The gluon building block $B^{(0)}_{n\perp}$ is only the leading term in the strong coupling expansion of the gauge invariant operator
\begin{equation}
  B^{\mu}_{n \perp} \equiv \frac{1}{g}  \lb W^{\dag}_n (\mathcal{P}_{\perp}^{\mu} -g A_{n \perp}^{\mu}) W_n \rb = {B}^{\mu, a(0)}_{n\perp} T^a + \mathcal{O}(g) \;.
\end{equation}
Written in terms of the effective Lagrangian, we have
\begin{equation}
  \mathcal{L}_{Q-G}^{(0)} (\psi,A^{\mu,a}_G)  = \sum_{\bmat{p},\bmat{q}_T}  \psi^{\dag}_{\bmat{p}+\bmat{q}_T} \lp  - g T^a v_{\mu} \rp \psi_{\bmat{p}}\; A^{\mu,a}_G   \;,
\end{equation}
where
\begin{equation}
  \label{eq:vector-G}
  A^{\mu,a}_G =  \frac{i}{2} g f^{abc} \frac{n^{\mu}}{\bmat{q}_{T}^2} \sum_{\ell} \lb \bar{n} \cdot \mathcal{P} \;(\bmat{B}^{b (0)}_{n\perp, \ell-\bmat{q}_T} \cdot \bmat{B}^{c (0)}_{n\perp,\ell}) \rb \;.
\end{equation}
Note that the form of the Lagrangian in terms of the effective Glauber field, $A^{a,\mu}_{G}$, remains the same as in Eqs.~(\ref{eq:L0-t-q}) and (\ref{eq:L0-NR}).  In the next-to-leading power expansion for the sum of all three diagrams we get
\begin{equation}
  \label{eq:t1-g-coll}
  t^{(1)}_{g-coll.} =-\frac{g^2}{2m} f^{abc} \lp 2m \xi^{\dag} \lbc (2 \bmat{r}_s+\bmat{q}_T) - i \bmat{q}_T \times \bmat{\sigma}  \rbc T^{c}  \xi \rp \cdot \bmat{n} \; \lb \frac{p_n^{-}}{\bmat{q}_T^{2}}\; \bmat{B}^{a (0)}_{n\perp, p_n} \cdot \bmat{B}^{b (0)}_{n \perp, p_n'} \rb \;.
\end{equation}
This gives
\begin{equation}
  \mathcal{L}_{Q-G}^{(1)} (\psi,A_{G}^{\mu,a}) =  g\sum_{\bmat{p},\bmat{q}_T} \psi^{\dag}_{\bmat{p} +\bmat{q}_T} \lp\frac{ 2  A_{G}^\bmat{n} (\bmat{n} \cdot \bmat{\mathcal{P}}) - i \lb ( \bmat{\mathcal{P}}_{\perp} \times \bmat{n}) A^{\bmat{n}}_{G}  \rb \cdot \bmat \sigma }{2m} \rp \psi_{\bmat{p}}\;,
\end{equation}
where the Glauber field, $A^{a,\mu}_G$  is given by Eq.~(\ref{eq:vector-G}). Comparing with the results for collinear quark sources we find that the Lagrangian in terms of the effective field $A_G^{\mu,a}$ is identical whichever collinear source (quark vs gluons) we are considering.
 
Repeating the same exercise for soft gluons, where we replace: $p_n \to p_s$ and $p_n' \to p_s'$ in Eq.(\ref{eq:g-diagrams}), we find
\begin{align}
  \label{eq:t1-g-soft}
  t^{(0)}_{g-soft} =& g^2 f^{abc}(2m \xi^{\dag} T^{c} \xi) \; \lb  \frac{2 p_s^{0}}{\bmat{q}^{2}}\; \bmat{B}^{a (0)}_{s, p_s} \cdot \bmat{B}^{b (0)}_{s, p_s'} \rb \; , \nn \\
  t^{(1)}_{g-soft} =& -i\frac{g^2}{2m} (2m \xi^{\dag}  \{ T^{a},T^{b} \} \xi)\lb   \bmat{B}^{a (0)}_{s, p_s} \cdot \bmat{B}^{b (0)}_{s, p_s'} \rb      
  + \frac{g^2}{2m} f^{abc} (2m \xi^{\dag} \bmat{\sigma} T^c  \xi) \cdot \lb   \bmat{B}^{a (0)}_{s, p_s} \times  \bmat{B}^{b (0)}_{s, p_s'} \rb           \nn \\
  &  - \frac{g^2}{2m \bmat{q}^{2}} f^{abc} \lp 2m \xi^{\dag} \lbc ( \bmat{r}_s+\bmat{r}_s') - i \bmat{q} \times \bmat{\sigma}  \rbc T^{c}  \xi \rp \cdot \lbc
  ( \bmat{p}_s +\bmat{p}_s')\; (\bmat{B}^{a (0)}_{s, p_s} \cdot \bmat{B}^{b (0)}_{s, p_s'} ) \nn \\
  &- 2 \bmat{B}^{b (0)}_{s, p_s'} (\bmat{p}_s' \cdot \bmat{B}^{a (0)}_{s, p_s}) - 2 \bmat{B}^{a (0)}_{s, p_s} (\bmat{p}_s \cdot \bmat{B}^{b (0)}_{s, p_s'})
  \rbc \;, 
\end{align}
where 
\begin{equation}
  \bmat{B}^{a,(0)}_{s,\ell}  \equiv \bmat{A}^{a}_{s,\ell}  - \bmat{p}_{s} \frac{ A^{0,a}_{s,\ell} }{p_{s}^0} \;.
\end{equation}
The soft gluon building block $B^{(0)}_{s}$ is only the leading term in the strong coupling expansion of the gauge invariant operator
\begin{equation}
  B^{\mu}_{s} \equiv \frac{1}{g}  \lb S^{\dag}_n (\mathcal{P}^{\mu} -g A_{s}^{\mu}) S_n \rb = {B}^{\mu, a(0)}_{s} T^a + \mathcal{O}(g) \;.
\end{equation}
In the forward scattering limit ($\bmat{q} \to 0$) this result can be further simplified and the corresponding Lagrangian, $\mathcal{L}_{Q-C}(\psi,A^{\mu,a}_C)$, in terms of the Coulomb field, $A_{C}^{\mu,a}$, can be written in the form of Eqs.~(\ref{eq:L0-NR}) and (\ref{eq:L1-NR}) where the effective Coulomb field in terms of the source soft gluon can be written as follows,
\begin{equation}
  \label{eq:vector-C}
  A^{\mu,a}_C =  f^{abc} \frac{i g}{2\; \bmat{q}^2} \sum_{\ell} \lbc
  \lb \mathcal{P}^{\mu} \;(\bmat{B}^{b (0)}_{s, \ell-\bmat{q}} \cdot \bmat{B}^{c (0)}_{s,\ell}) \rb
  - 2 (\bmat{B}^{c (0)}_{s,\ell} \cdot \lb \bmat{\mathcal{P}} ) B^{\mu,b (0)}_{s, \ell-\bmat{q}}  \rb
  -2 (\bmat{B}^{b (0)}_{s,\ell -\bmat{q}} \cdot \lb \bmat{\mathcal{P}} ) B^{\mu,c (0)}_{s, \ell}  \rb
  \rbc\;.
\end{equation}
Note that from the equation of motion, $v \cdot B^{(0)} = 0$, the last two terms in Eq.~(\ref{eq:vector-C}) will not contribute to the leading Lagrangian, $\mathcal{L}^{(0)}_{Q-C}$.
\subsection{Comparison with the literature}
\label{sec:comp}
The interaction of heavy quarks with soft fermions and gluons was studied in the framework of vNRQCD in Refs.~\cite{Luke:1999kz, Rothstein:2018dzq}. Here,  we are interested in the case where the fields are sourced by partons originating from a quark-gluon plasma (or some other medium), but the formalism (non-relativistic expansion) up-to the effective coupling remains the same. Therefore, we test our approach be comparing our result in Eq.~(\ref{eq:soft-t}) for soft fermion sources with those of Eqs.~(2.9), (2.10), and (3.11) of Ref.~\cite{Rothstein:2018dzq} and find that the two agree. Note the overall $i$ factor from expanding the action, also in our notation $\bmat{q} = \bmat{r}'_s -\bmat{r}_s$. For interactions of the heavy quarks with soft gluons,  one should  then compare our Eq.~(\ref{eq:t1-g-soft}) with Eqs.~(3.6), (3.7), and (3.11) of Ref.~\cite{Rothstein:2018dzq}. Again, the two results are in agreement and we note also the factor of 1/2 introduced at the level of the Lagrangian for the symmetry of exchanging the two soft gluons.

The interactions of heavy quarks with collinear partons were studied in the context of SCET$_{\rm G}$ in Ref.~\cite{Ovanesyan:2011xy}, where only the leading Lagrangian, $\mathcal{L}_{Q-G}^{(0)}$, was investigated. For interactions with collinear quarks our result in Eq.~(\ref{eq:t0-q-coll}) agrees with the equivalent result in Eq.~(4.14) of Ref.~\cite{Ovanesyan:2011xy}.  In contrast, for interactions with collinear gluons our results in Eq.~(\ref{eq:t0-g-coll}) disagree with the corresponding of Ref.~\cite{Ovanesyan:2011xy}. The disagreement originates from the fact that in~\cite{Ovanesyan:2011xy} the authors consider only the first of the three diagrams and assume the replacement $A^{\mu} \to B^{\mu}_{n\perp}$.
For forward scattering processes on the medium quasiparticles to lowest non-trivial order, this is the dominant diagram and the gauge invariance of the splitting kernels was checked explicitly by comparing three different gauges: covariant,  lightcone, and hybrid. For the general cause, however, we expect that this will not be true. Here, we establish gauge invariance most generally  at the level of the matching procedure.  Furthermore, to our knowledge the results for $\mathcal{L}_{Q-G}^{(1)}$ are new both for collinear quarks and gluons. 


\section{Conclusions}
\label{sec:conclusions}

In recent years, different phenomenological  approaches have been proposed to describe  the modification of the production cross sections of moderate and high transverse momentum quarkonia. Theoretical guidance on the relative significance of the various nuclear effects in the currently accessible transverse momentum range can be very useful. In this paper we used the leading power factorization limit of NRQCD, along with recent extractions of the LDMEs,  to implement the energy loss approach to quarkonium production. We calculated the $J/\psi$ and $\psi(2S)$ suppression in the $p_T =10-40$ GeV range and compared the theoretical predictions  to experimental measurements from ATLAS  and CMS collaborations at $\sqrt{s} = 5.02$ GeV for Pb-Pb collisions. We found that theoretical predictions  overestimate of the $J/\psi$ suppression for both 0-10\% and 0-80\% central collisions and the discrepancies persist even after taking the effective coupling to be smaller than  traditionally used for in-medium jet propagation.  Most importantly, comparing the double radio $R_{\text{AA}}[\psi(2S)]/ R_{\text{AA}}[J/\psi]$  to data, we also find a disagreement  that cannot be resolved  within the energy loss model. Wwhile the data show that suppression of exited states is clearly larger by more than  a factor of two, the theoretical prediction yields a distinctly opposite trend, suppression of the $J/\psi$ is slightly larger.

The strong tension between experimental data and theoretical predictions suggests that the energy loss assumption for production and propagation of quarkonium states  in medium needs to be revisited. As a formal step in that direction, we introduced a modified theory of non-relativistic QCD that accounts for the interactions of heavy quarks and antiquarks with the medium through soft-virtual gluon exchanges. We refer to the resulting effective theory as NRQCD$_{\rm G}$ and  considered three types of medium sources for the virtual gluons: static, soft, and collinear. For static and soft sources we identified the Coulomb region, $q_C^{\mu} \sim (\lambda^2, \lambda, \lambda, \lambda)$, to be the most relevant. On the other hand, for collinear sources the leading contributions come from the Glauber region, $q_G^{\mu} \sim (\lambda^2, \lambda, \lambda, \lambda^2)$. 

We derived the NRQCD$_{\rm G}$ leading and sub-leading Lagrangians for a single virtual gluon exchange.  To accomplish this task, we used three different approaches: i) the background field method, ii) a matching (with QCD) procedure, and iii) a hybrid method.  Although we found that applying the background field method requires caution in the order of shifting the fields and applying power-counting (as discussed in Section~\ref{sec:bfa} and Appendix~\ref{app:last}), all three methods give the same Lagrangian which serves as a non-trivial test of our derivation. A natural extension of this work will be to also extract  the double virtual gluon interactions. This can be achieved with minimal effort in the background field method, as described in Appendix~\ref{app:last}, but a consistency check through one of the other two approaches is advisable. We have outlined the process of such derivation in the hybrid model below Eq.~(\ref{eq:2A-diagrams}).

As we focused on the formal aspects of  of NRQCD$_{\rm G}$,  phenomenological applications to various topics of interest  are left for the future. In particular,  would be interesting to investigate using the EFT derived in this work the modification of the heavy quark-antiquark potential due to medium interactions, which in the vacuum is   Coulomb-like. In addition,  interactions with the medium could induce radial excitations which will likely cause transitions from one quarkonium state to another.  Medium-induced transitions from and to exited states might modify the observed relative suppression rates. Moreover, it is interesting to entertain the possibility of using the terms from the matching procedure to investigate the effect of Glauber gluons in quarkonium production and decay factorization theorems in the vacuum.


\begin{acknowledgments}
We would like to thank Christopher Lee  and Rishi Sharma for useful discussions during the course of this project. This work was supported by the U.S. Department of Energy, Office of Science Contract DE-AC52-06NA25396,   the LANL LDRD Program under Contract 20190033ER, and the DOE's Early Career Program.
\end{acknowledgments}

\appendix
\section{The background field approach revised}
\label{app:last}
As commented below Eq.~(\ref{eq:L1-NR}), the background field approach that was implemented in Section~\ref{sec:bfa} yields different results compared to the non-relativistic limit of QCD. The discrepancy can be traced to the level of distinction of soft and ultra-soft modes. For one to arrive to the form and power-counting of the various terms in the Lagrangian, one has to assume  scaling of the gluon filed $A^{\mu,a}_U$ and its momenta, which in this case is ultra-soft. Therefore, shifting the field to include the Glauber or Coulomb gluons which have components of their momenta scaling as soft rather as ultra-soft, results in missing various terms. It is, thus,  important to start from a point at which the soft and ultra-soft distinction is not yet made. Conveniently, this is the standard NRQCD Lagrangian. In particular, we are considering Eqs.~(2.4) and (2.5) of Ref.~\cite{Bodwin:1994jh}. 

In order to extract the Glauber and Coulomb insertions from the NRQCD Lagrangian, but yet formulate the final result in the label momentum notation, we will perform the following replacements
\begin{align}
  \psi (x) &\rightarrow  \sum_{\bmat{p}} \psi_{\bmat{p}} (x) \;, \nn  \\ 
  iD_{\mu} &\rightarrow \mathcal{P}_{\mu} + i \partial_{\mu} - g (A_U^{\mu}+A_{G/C}^{\mu}) \;,
\end{align}
where it is understood that after the replacement the partial derivatives act only on the conjugate of ultra-soft momenta. The four-momentum version of the label momentum operator is defined as  $\mathcal{P}_{\mu} = (0,-\bmat{\mathcal{P}})$. In order to perform the analysis in an organized manner is important to establish the power-counting of the various operator that appear in the Lagrangian. We will conciser each source separately. We start with the collinear source.
\begin{align}
  iD_{t} &= \underbrace{i \partial_t - g A_{U}^{0} -g A_{G}^{0}}_{\textstyle{\sim \lambda^2}} \;, \nn \\ 
  i\bmat{D}& = \underbrace{\bmat{\mathcal{P}}}_{\textstyle{\sim \lambda}} - (\underbrace{ i \bmat{\partial} + g \bmat{A}_{U} + g \bmat{n} A_{G}^{\bmat{n}}}_{\textstyle{\sim \lambda^2}})
  + \mathcal{O}(\lambda^{3}) \;, \nn \\ 
  \bmat{E} &= \partial_t (\bmat{A}_U+\bmat{A}_G) + (\bmat{\partial} + i \bmat{\mathcal{P}}) (A_{U}^{0} + A_{G}^{0}) + g T^{c} f^{cba}  (A_{U}^{0} + A_{G}^{0})^{b}
  (\bmat{A}_U+\bmat{A}_G)^{a} \nn \\
  &= \underbrace{  i \bmat{\mathcal{P}}_{\perp} A_{G}^{0} }_{\textstyle{\sim \lambda^3}} + \mathcal{O}(\lambda^4)  \;,\nn \\ 
  \bmat{B} &= - (\bmat{\partial} + i \bmat{\mathcal{P}})\times (\bmat{A}_U+\bmat{A}_G) + \frac{g}{2} T^{c} f^{cba}   (\bmat{A}_U+\bmat{A}_G)^{b} (\bmat{A}_U+\bmat{A}_G)^{a} \nn\\
  &= -\underbrace{  (i \bmat{\mathcal{P}}_{\perp} \times\bmat{n})\; A^{\bmat{n}}_{G} }_{\textstyle{\sim \lambda^3}} + \mathcal{O}(\lambda^4) \;.
\end{align}
We now have all the ingredients to expand the Lagrangian up to $\mathcal{O}(\lambda^{3})$. \footnote{ This does not include the power-counting of the heavy quark filed $\psi_{\bmat{p}}(x) \sim \lambda^{3/2}$ since it appear for all the terms we are considering here.} Collecting all the terms that do not involve the  field $A_{G}$ will give us the heavy quark part of the vNRQCD Lagrangian. For $\mathcal{L}_{Q-G}$ we need to collect all the terms that contain at least one power of $A_{G}$. We, thus, get:
\begin{equation}
(collinear)\;\;\;\;\; \mathcal{L}_{Q-G} (\psi,A_{G}^{\mu,a})  = g \sum_{\bmat{p},\bmat{q}_T}\psi^{\dag}_{\bmat{p}+\bmat{q}_T} \lp -  A^{0}_{G} +  \frac{ 2  A_{G}^\bmat{n} (\bmat{n} \cdot \bmat{\mathcal{P}}) - i \lb ( \bmat{\mathcal{P}}_{\perp} \times \bmat{n}) A^{\bmat{n}}_{G}  \rb \cdot \bmat \sigma }{2m} \rp \psi_{\bmat{p}} \;.
\end{equation}
This result is exactly what we obtain when we sum the leading and sub-leading terms, i.e. $\mathcal{L}^{(0)}_{Q-G}+ \mathcal{L}^{(1)}_{Q-G}$ from Eqs.~(\ref{eq:L0-NR}) and Eqs.~(\ref{eq:L1-NR}). If we now instead consider a static source, then the scaling of the same operators is as follows,
\begin{align}
  iD_{t} &=  \underbrace{-g A_{C}^{0}}_{\textstyle{\sim \lambda}} + (\underbrace{i \partial_t - g A_{U}^{0}}_{\textstyle{\sim \lambda^2}}) \;,  \nn \\ 
  i\bmat{D}& = \underbrace{\bmat{\mathcal{P}}}_{\textstyle{\sim \lambda}} - (\underbrace{ i \bmat{\partial} + g \bmat{A}_{U} + g \bmat{A}_{G} }_{\textstyle{\sim \lambda^2}}) \;, 
  \nn \\ 
  \bmat{E} &= \partial_t (\bmat{A}_U+\bmat{A}_C) + (\bmat{\partial} + i \bmat{\mathcal{P}}) (A_{U}^{0} + A_{C}^{0}) + g T^{c} f^{cba}  (A_{U}^{0} + A_{C}^{0})^{b}
  (\bmat{A}_U+\bmat{A}_C)^{a} \nn \\
  & = \underbrace{  i \bmat{\mathcal{P}} A_{C}^{0} }_{\textstyle{\sim \lambda^2}} + \mathcal{O}(\lambda^3) \;,  \nn \\
  \bmat{B} &= - (\bmat{\partial} + i \bmat{\mathcal{P}})\times (\bmat{A}_U+\bmat{A}_C) + \frac{g}{2} T^{c} f^{cba}   (\bmat{A}_U+\bmat{A}_C)^{b} (\bmat{A}_U+\bmat{A}_C)^{a}
  = -\underbrace{  i \bmat{\mathcal{P}} \times\bmat{A}_{C} }_{\textstyle{\sim \lambda^3}} + \mathcal{O}(\lambda^4) \;.
\end{align}
Collecting all the terms that involve the field $A_{C}$ we get:
\begin{equation}
  (static)\;\;\;\;\; \mathcal{L}_{Q-C} (\psi,A_{C}^{\mu,a})  = g \sum_{\bmat{p},\bmat{q}_T}\psi^{\dag}_{\bmat{p}+\bmat{q}_T} \lp -  A^{0}_{C} +  \mathcal{O}(\lambda^3)\rp \psi_{\bmat{p}} \;.
\end{equation}
Again, this is exactly what we  can derive by summing the leading and sub-leading terms, i.e. $\mathcal{L}^{(0)}_{Q-C}+ \mathcal{L}^{(1)}_{Q-C}$ from Eqs.~(\ref{eq:L0-NR}) and Eqs.~(\ref{eq:L1-NR}). We are now ready for the final derivation of this appendix. We implement the same analysis as above for a soft source. Then we have
\begin{align}
  iD_{t} &=  \underbrace{-g A_{C}^{0}}_{\textstyle{\sim \lambda}} + (\underbrace{i \partial_t - g A_{U}^{0}}_{\textstyle{\sim \lambda^2}}) \;, \nn \\ 
  i\bmat{D}& = (\underbrace{\bmat{\mathcal{P}} - g \bmat{A}_{C}}_{\textstyle{\sim \lambda}}) - (\underbrace{ i \bmat{\partial} + g \bmat{A}_{U}}_{\textstyle{\sim \lambda^2}}) \;,
  \nn \\ 
  \bmat{E} &= \partial_t (\bmat{A}_U+\bmat{A}_C) + (\bmat{\partial} + i \bmat{\mathcal{P}}) (A_{U}^{0} + A_{C}^{0}) + g T^{c} f^{cba}  (A_{U}^{0} + A_{C}^{0})^{b}
  (\bmat{A}_U+\bmat{A}_C)^{a} \nn \\&
  = (\underbrace{  i \bmat{\mathcal{P}} A_{C}^{0} + i g [ \bmat{A}_C,  A_{C}^{0}]}_{\textstyle{\sim \lambda^2}}) + \mathcal{O}(\lambda^3)  \;, \nn \\ 
  \bmat{B} &= - (\bmat{\partial} + i \bmat{\mathcal{P}})\times (\bmat{A}_U+\bmat{A}_C) + \frac{g}{2} T^{c} f^{cba}   (\bmat{A}_U+\bmat{A}_C)^{b} \times (\bmat{A}_U+\bmat{A}_C)^{a} 
  \nn \\  &= - i \underbrace{  ( \bmat{\mathcal{P}}  + g \bmat{A}_C ) \times \bmat{A}_{C}}_{\textstyle{\sim \lambda^2}} + \mathcal{O}(\lambda^3) \; .
\end{align}
Collecting all the terms that involve the field $A_{C}$ we get:
\begin{multline}
 (soft)\\ \mathcal{L}_{Q-C} (\psi,A_{C}^{\mu,a})  = g \sum_{\bmat{p},\bmat{q}_T}\psi^{\dag}_{\bmat{p}+\bmat{q}_T} \lp -  A^{0}_{C} + \frac{ 2  \bmat{A}_{C} \cdot \bmat{\mathcal{P}} + [\bmat{\mathcal{P}} \cdot  \bmat{A}_{C} ] - i \lb  \bmat{\mathcal{P}} \times \bmat{A}_{C} \rb \cdot \bmat \sigma }{2m}\rp \psi_{\bmat{p}} + \mathcal{O}(g^2) \, .
\end{multline}
The sum of the leading and sub-leading terms, i.e. $\mathcal{L}^{(0)}_{Q-C}+ \mathcal{L}^{(1)}_{Q-C}$ from Eqs.~(\ref{eq:L0-NR}) and Eqs.~(\ref{eq:L1-NR}), is identical to this result. The order $\mathcal{O}(g^2)$ terms we excluded in the above equation are,
\begin{equation}
  - g^2 \sum_{\bmat{p},\bmat{q}_T, \bmat{q}'_T}\psi^{\dag}_{\bmat{p}+\bmat{q}_T  +\bmat{q}'_T} \lp  \frac{\bmat{A}_{C}^2 + i \bmat{A}_{C}  \times \bmat{A}_{C} }{2m} +  \mathcal{O}(\lambda^3)\rp \psi_{\bmat{p}} \;.
  \end{equation}
 As mentioned earlier these terms can be reproduced in the hybrid method or within the matching procedure by evaluating Eq.~(\ref{eq:2A-diagrams}). We do not pursue this task here.

It is important to mention that in this section we only study the tree-level result for the NRQCD$_{\rm G}$ Lagrangian. The coefficients for the various terms in the Lagrangian take loop corrections and the coefficients can be written as an expansion in the strong coupling. Logarithmic enchantments in the perturbative expansion of the coefficients  need to be resumed through renormalization group equations (RGEs). An important question is if the perturbative expansion for these coefficients remains the same as in NRQCD. Using the background filed approach the coefficients, by construction, do not change. This is, obviously, a very nontrivial statement if using a matching approach.  Further studies of this  issue might be an  important
task for the future.

\cleardoublepage
\bibliographystyle{JHEP}
\normalbaselines 
\bibliography{notes} 

\providecommand{\href}[2]{#2}\begingroup\raggedright\begin{thebibliography}{10}

\bibitem{BraunMunzinger:2007zz}
P.~Braun-Munzinger and J.~Stachel, \emph{{The quest for the quark-gluon
  plasma}}, \href{http://dx.doi.org/10.1038/nature06080}{\emph{Nature} {\bf
  448} (2007) 302--309}.

\bibitem{Matsui:1986dk}
T.~Matsui and H.~Satz, \emph{{$J/\psi$ Suppression by Quark-Gluon Plasma
  Formation}},
  \href{http://dx.doi.org/10.1016/0370-2693(86)91404-8}{\emph{Phys. Lett.} {\bf
  B178} (1986) 416--422}.

\bibitem{Mocsy:2007jz}
A.~Mocsy and P.~Petreczky, \emph{{Color screening melts quarkonium}},
  \href{http://dx.doi.org/10.1103/PhysRevLett.99.211602}{\emph{Phys. Rev.
  Lett.} {\bf 99} (2007) 211602}, [\href{http://arxiv.org/abs/0706.2183}{{\tt
  0706.2183}}].

\bibitem{Krouppa:2015yoa}
B.~Krouppa, R.~Ryblewski and M.~Strickland, \emph{{Bottomonia suppression in
  2.76 TeV Pb-Pb collisions}},
  \href{http://dx.doi.org/10.1103/PhysRevC.92.061901}{\emph{Phys. Rev.} {\bf
  C92} (2015) 061901}, [\href{http://arxiv.org/abs/1507.03951}{{\tt
  1507.03951}}].

\bibitem{Hoelck:2016tqf}
J.~Hoelck, F.~Nendzig and G.~Wolschin, \emph{{In-medium $\Upsilon$ suppression
  and feed-down in UU and PbPb collisions}},
  \href{http://dx.doi.org/10.1103/PhysRevC.95.024905}{\emph{Phys. Rev.} {\bf
  C95} (2017) 024905}, [\href{http://arxiv.org/abs/1602.00019}{{\tt
  1602.00019}}].

\bibitem{Du:2017qkv}
X.~Du, R.~Rapp and M.~He, \emph{{Color Screening and Regeneration of Bottomonia
  in High-Energy Heavy-Ion Collisions}},
  \href{http://dx.doi.org/10.1103/PhysRevC.96.054901}{\emph{Phys. Rev.} {\bf
  C96} (2017) 054901}, [\href{http://arxiv.org/abs/1706.08670}{{\tt
  1706.08670}}].

\bibitem{Aronson:2017ymv}
S.~Aronson, E.~Borras, B.~Odegard, R.~Sharma and I.~Vitev, \emph{{Collisional
  and thermal dissociation of $J/\psi$ and $\Upsilon$ states at the LHC}},
  \href{http://dx.doi.org/10.1016/j.physletb.2018.01.038}{\emph{Phys. Lett.}
  {\bf B778} (2018) 384--391}, [\href{http://arxiv.org/abs/1709.02372}{{\tt
  1709.02372}}].

\bibitem{Jamal:2018mog}
M.~Y. Jamal, I.~Nilima, V.~Chandra and V.~K. Agotiya, \emph{{Dissociation of
  heavy quarkonia in an anisotropic hot QCD medium in a quasiparticle model}},
  \href{http://dx.doi.org/10.1103/PhysRevD.97.094033}{\emph{Phys. Rev.} {\bf
  D97} (2018) 094033}, [\href{http://arxiv.org/abs/1805.04763}{{\tt
  1805.04763}}].

\bibitem{Laine:2006ns}
M.~Laine, O.~Philipsen, P.~Romatschke and M.~Tassler, \emph{{Real-time static
  potential in hot QCD}},
  \href{http://dx.doi.org/10.1088/1126-6708/2007/03/054}{\emph{JHEP} {\bf 03}
  (2007) 054}, [\href{http://arxiv.org/abs/hep-ph/0611300}{{\tt
  hep-ph/0611300}}].

\bibitem{Brambilla:2008cx}
N.~Brambilla, J.~Ghiglieri, A.~Vairo and P.~Petreczky, \emph{{Static
  quark-antiquark pairs at finite temperature}},
  \href{http://dx.doi.org/10.1103/PhysRevD.78.014017}{\emph{Phys. Rev.} {\bf
  D78} (2008) 014017}, [\href{http://arxiv.org/abs/0804.0993}{{\tt
  0804.0993}}].

\bibitem{Burnier:2015tda}
Y.~Burnier, O.~Kaczmarek and A.~Rothkopf, \emph{{Quarkonium at finite
  temperature: Towards realistic phenomenology from first principles}},
  \href{http://dx.doi.org/10.1007/JHEP12(2015)101}{\emph{JHEP} {\bf 12} (2015)
  101}, [\href{http://arxiv.org/abs/1509.07366}{{\tt 1509.07366}}].

\bibitem{Riek:2010fk}
F.~Riek and R.~Rapp, \emph{{Quarkonia and Heavy-Quark Relaxation Times in the
  Quark-Gluon Plasma}},
  \href{http://dx.doi.org/10.1103/PhysRevC.82.035201}{\emph{Phys. Rev.} {\bf
  C82} (2010) 035201}, [\href{http://arxiv.org/abs/1005.0769}{{\tt
  1005.0769}}].

\bibitem{Sharma:2012dy}
R.~Sharma and I.~Vitev, \emph{{High transverse momentum quarkonium production
  and dissociation in heavy ion collisions}},
  \href{http://dx.doi.org/10.1103/PhysRevC.87.044905}{\emph{Phys. Rev.} {\bf
  C87} (2013) 044905}, [\href{http://arxiv.org/abs/1203.0329}{{\tt
  1203.0329}}].

\bibitem{Ferreiro:2014bia}
E.~G. Ferreiro, \emph{{Excited charmonium suppression in proton–nucleus
  collisions as a consequence of comovers}},
  \href{http://dx.doi.org/10.1016/j.physletb.2015.07.066}{\emph{Phys. Lett.}
  {\bf B749} (2015) 98--103}, [\href{http://arxiv.org/abs/1411.0549}{{\tt
  1411.0549}}].

\bibitem{Akamatsu:2011se}
Y.~Akamatsu and A.~Rothkopf, \emph{{Stochastic potential and quantum
  decoherence of heavy quarkonium in the quark-gluon plasma}},
  \href{http://dx.doi.org/10.1103/PhysRevD.85.105011}{\emph{Phys. Rev.} {\bf
  D85} (2012) 105011}, [\href{http://arxiv.org/abs/1110.1203}{{\tt
  1110.1203}}].

\bibitem{Brambilla:2016wgg}
N.~Brambilla, M.~A. Escobedo, J.~Soto and A.~Vairo, \emph{{Quarkonium
  suppression in heavy-ion collisions: an open quantum system approach}},
  \href{http://dx.doi.org/10.1103/PhysRevD.96.034021}{\emph{Phys. Rev.} {\bf
  D96} (2017) 034021}, [\href{http://arxiv.org/abs/1612.07248}{{\tt
  1612.07248}}].

\bibitem{Kajimoto:2017rel}
S.~Kajimoto, Y.~Akamatsu, M.~Asakawa and A.~Rothkopf, \emph{{Dynamical
  dissociation of quarkonia by wave function decoherence}},
  \href{http://dx.doi.org/10.1103/PhysRevD.97.014003}{\emph{Phys. Rev.} {\bf
  D97} (2018) 014003}, [\href{http://arxiv.org/abs/1705.03365}{{\tt
  1705.03365}}].

\bibitem{Yao:2018nmy}
X.~Yao and T.~Mehen, \emph{{Quarkonium in-Medium Transport Equation Derived
  from First Principles}},  \href{http://arxiv.org/abs/1811.07027}{{\tt
  1811.07027}}.

\bibitem{Chatrchyan:2013nza}
{\scshape CMS} collaboration, S.~Chatrchyan et~al., \emph{{Event activity
  dependence of Y(nS) production in $\sqrt{s_{NN}}$=5.02 TeV pPb and
  $\sqrt{s}$=2.76 TeV pp collisions}},
  \href{http://dx.doi.org/10.1007/JHEP04(2014)103}{\emph{JHEP} {\bf 04} (2014)
  103}, [\href{http://arxiv.org/abs/1312.6300}{{\tt 1312.6300}}].

\bibitem{Ferreiro:2018wbd}
E.~G. Ferreiro and J.-P. Lansberg, \emph{{Is bottomonium suppression in
  proton-nucleus and nucleus-nucleus collisions at LHC energies due to the same
  effects?}}, \href{http://dx.doi.org/10.1007/JHEP03(2019)063,
  10.1007/JHEP10(2018)094}{\emph{JHEP} {\bf 10} (2018) 094},
  [\href{http://arxiv.org/abs/1804.04474}{{\tt 1804.04474}}].

\bibitem{Adare:2013ezl}
{\scshape PHENIX} collaboration, A.~Adare et~al., \emph{{Nuclear Modification
  of $??, ?_c$, and J/? Production in d+Au Collisions at
  $\sqrt{s_{NN}}$=200??GeV}},
  \href{http://dx.doi.org/10.1103/PhysRevLett.111.202301}{\emph{Phys. Rev.
  Lett.} {\bf 111} (2013) 202301}, [\href{http://arxiv.org/abs/1305.5516}{{\tt
  1305.5516}}].

\bibitem{Adamczyk:2013poh}
{\scshape STAR} collaboration, L.~Adamczyk et~al., \emph{{Suppression of
  $\Upsilon$ production in d+Au and Au+Au collisions at $\sqrt{s_{NN}}$=200
  GeV}}, \href{http://dx.doi.org/10.1016/j.physletb.2014.06.028,
  10.1016/j.physletb.2015.01.046}{\emph{Phys. Lett.} {\bf B735} (2014)
  127--137}, [\href{http://arxiv.org/abs/1312.3675}{{\tt 1312.3675}}].

\bibitem{Almeida:2014uva}
L.~G. Almeida, S.~D. Ellis, C.~Lee, G.~Sterman, I.~Sung and J.~R. Walsh,
  \emph{{Comparing and counting logs in direct and effective methods of QCD
  resummation}}, \href{http://dx.doi.org/10.1007/JHEP04(2014)174}{\emph{JHEP}
  {\bf 04} (2014) 174}, [\href{http://arxiv.org/abs/1401.4460}{{\tt
  1401.4460}}].

\bibitem{Bodwin:1994jh}
G.~T. Bodwin, E.~Braaten and G.~P. Lepage, \emph{{Rigorous QCD analysis of
  inclusive annihilation and production of heavy quarkonium}},
  \href{http://dx.doi.org/10.1103/PhysRevD.55.5853,
  10.1103/PhysRevD.51.1125}{\emph{Phys. Rev.} {\bf D51} (1995) 1125--1171},
  [\href{http://arxiv.org/abs/hep-ph/9407339}{{\tt hep-ph/9407339}}].

\bibitem{Lansberg:2019adr}
J.-P. Lansberg, \emph{{New Observables in Inclusive Production of Quarkonia}},
  \href{http://arxiv.org/abs/1903.09185}{{\tt 1903.09185}}.

\bibitem{Bauer:2000ew}
C.~W. Bauer, S.~Fleming and M.~E. Luke, \emph{{Summing Sudakov logarithms in B
  ---> X(s gamma) in effective field theory}},
  \href{http://dx.doi.org/10.1103/PhysRevD.63.014006}{\emph{Phys. Rev.} {\bf
  D63} (2000) 014006}, [\href{http://arxiv.org/abs/hep-ph/0005275}{{\tt
  hep-ph/0005275}}].

\bibitem{Bauer:2000yr}
C.~W. Bauer, S.~Fleming, D.~Pirjol and I.~W. Stewart, \emph{{An Effective field
  theory for collinear and soft gluons: Heavy to light decays}},
  \href{http://dx.doi.org/10.1103/PhysRevD.63.114020}{\emph{Phys. Rev.} {\bf
  D63} (2001) 114020}, [\href{http://arxiv.org/abs/hep-ph/0011336}{{\tt
  hep-ph/0011336}}].

\bibitem{Bauer:2001ct}
C.~W. Bauer and I.~W. Stewart, \emph{{Invariant operators in collinear
  effective theory}},
  \href{http://dx.doi.org/10.1016/S0370-2693(01)00902-9}{\emph{Phys. Lett.}
  {\bf B516} (2001) 134--142}, [\href{http://arxiv.org/abs/hep-ph/0107001}{{\tt
  hep-ph/0107001}}].

\bibitem{Bauer:2001yt}
C.~W. Bauer, D.~Pirjol and I.~W. Stewart, \emph{{Soft collinear factorization
  in effective field theory}},
  \href{http://dx.doi.org/10.1103/PhysRevD.65.054022}{\emph{Phys. Rev.} {\bf
  D65} (2002) 054022}, [\href{http://arxiv.org/abs/hep-ph/0109045}{{\tt
  hep-ph/0109045}}].

\bibitem{Baumgart:2014upa}
M.~Baumgart, A.~K. Leibovich, T.~Mehen and I.~Z. Rothstein, \emph{{Probing
  Quarkonium Production Mechanisms with Jet Substructure}},
  \href{http://dx.doi.org/10.1007/JHEP11(2014)003}{\emph{JHEP} {\bf 11} (2014)
  003}, [\href{http://arxiv.org/abs/1406.2295}{{\tt 1406.2295}}].

\bibitem{Bain:2016rrv}
R.~Bain, Y.~Makris and T.~Mehen, \emph{{Transverse Momentum Dependent
  Fragmenting Jet Functions with Applications to Quarkonium Production}},
  \href{http://dx.doi.org/10.1007/JHEP11(2016)144}{\emph{JHEP} {\bf 11} (2016)
  144}, [\href{http://arxiv.org/abs/1610.06508}{{\tt 1610.06508}}].

\bibitem{Idilbi:2008vm}
A.~Idilbi and A.~Majumder, \emph{{Extending Soft-Collinear-Effective-Theory to
  describe hard jets in dense QCD media}},
  \href{http://dx.doi.org/10.1103/PhysRevD.80.054022}{\emph{Phys. Rev.} {\bf
  D80} (2009) 054022}, [\href{http://arxiv.org/abs/0808.1087}{{\tt
  0808.1087}}].

\bibitem{Ovanesyan:2011xy}
G.~Ovanesyan and I.~Vitev, \emph{{An effective theory for jet propagation in
  dense QCD matter: jet broadening and medium-induced bremsstrahlung}},
  \href{http://dx.doi.org/10.1007/JHEP06(2011)080}{\emph{JHEP} {\bf 06} (2011)
  080}, [\href{http://arxiv.org/abs/1103.1074}{{\tt 1103.1074}}].

\bibitem{Kang:2016ofv}
Z.-B. Kang, F.~Ringer and I.~Vitev, \emph{{Effective field theory approach to
  open heavy flavor production in heavy-ion collisions}},
  \href{http://dx.doi.org/10.1007/JHEP03(2017)146}{\emph{JHEP} {\bf 03} (2017)
  146}, [\href{http://arxiv.org/abs/1610.02043}{{\tt 1610.02043}}].

\bibitem{Kang:2014xsa}
Z.-B. Kang, R.~Lashof-Regas, G.~Ovanesyan, P.~Saad and I.~Vitev, \emph{{Jet
  quenching phenomenology from soft-collinear effective theory with Glauber
  gluons}}, \href{http://dx.doi.org/10.1103/PhysRevLett.114.092002}{\emph{Phys.
  Rev. Lett.} {\bf 114} (2015) 092002},
  [\href{http://arxiv.org/abs/1405.2612}{{\tt 1405.2612}}].

\bibitem{Chien:2015vja}
Y.-T. Chien, A.~Emerman, Z.-B. Kang, G.~Ovanesyan and I.~Vitev, \emph{{Jet
  Quenching from QCD Evolution}},
  \href{http://dx.doi.org/10.1103/PhysRevD.93.074030}{\emph{Phys. Rev.} {\bf
  D93} (2016) 074030}, [\href{http://arxiv.org/abs/1509.02936}{{\tt
  1509.02936}}].

\bibitem{Spousta:2016agr}
M.~Spousta, \emph{{On similarity of jet quenching and charmonia suppression}},
  \href{http://dx.doi.org/10.1016/j.physletb.2017.01.041}{\emph{Phys. Lett.}
  {\bf B767} (2017) 10--15}, [\href{http://arxiv.org/abs/1606.00903}{{\tt
  1606.00903}}].

\bibitem{Arleo:2017ntr}
F.~Arleo, \emph{{Quenching of Hadron Spectra in Heavy Ion Collisions at the
  LHC}}, \href{http://dx.doi.org/10.1103/PhysRevLett.119.062302}{\emph{Phys.
  Rev. Lett.} {\bf 119} (2017) 062302},
  [\href{http://arxiv.org/abs/1703.10852}{{\tt 1703.10852}}].

\bibitem{Khachatryan:2016ypw}
{\scshape CMS} collaboration, V.~Khachatryan et~al., \emph{{Suppression and
  azimuthal anisotropy of prompt and nonprompt ${\mathrm{J}}/\psi $ production
  in PbPb collisions at $\sqrt{{s_{_{\text {NN}}}}} =2.76$ $\,\mathrm{TeV}$}},
  \href{http://dx.doi.org/10.1140/epjc/s10052-017-4781-1}{\emph{Eur. Phys. J.}
  {\bf C77} (2017) 252}, [\href{http://arxiv.org/abs/1610.00613}{{\tt
  1610.00613}}].

\bibitem{Aaboud:2018quy}
{\scshape ATLAS} collaboration, M.~Aaboud et~al., \emph{{Prompt and non-prompt
  $J/\psi $ and $\psi (2\mathrm {S})$ suppression at high transverse momentum
  in $5.02~\mathrm {TeV}$ Pb+Pb collisions with the ATLAS experiment}},
  \href{http://dx.doi.org/10.1140/epjc/s10052-018-6219-9}{\emph{Eur. Phys. J.}
  {\bf C78} (2018) 762}, [\href{http://arxiv.org/abs/1805.04077}{{\tt
  1805.04077}}].

\bibitem{Luke:1999kz}
M.~E. Luke, A.~V. Manohar and I.~Z. Rothstein, \emph{{Renormalization group
  scaling in nonrelativistic QCD}},
  \href{http://dx.doi.org/10.1103/PhysRevD.61.074025}{\emph{Phys. Rev.} {\bf
  D61} (2000) 074025}, [\href{http://arxiv.org/abs/hep-ph/9910209}{{\tt
  hep-ph/9910209}}].

\bibitem{Rothstein:2018dzq}
I.~Z. Rothstein, P.~Shrivastava and I.~W. Stewart, \emph{{Manifestly Soft Gauge
  Invariant Formulation of vNRQCD}},
  \href{http://arxiv.org/abs/1806.07398}{{\tt 1806.07398}}.

\bibitem{Cho:1995vh}
P.~L. Cho and A.~K. Leibovich, \emph{{Color octet quarkonia production}},
  \href{http://dx.doi.org/10.1103/PhysRevD.53.150}{\emph{Phys. Rev.} {\bf D53}
  (1996) 150--162}, [\href{http://arxiv.org/abs/hep-ph/9505329}{{\tt
  hep-ph/9505329}}].

\bibitem{Cho:1995ce}
P.~L. Cho and A.~K. Leibovich, \emph{{Color octet quarkonia production. 2.}},
  \href{http://dx.doi.org/10.1103/PhysRevD.53.6203}{\emph{Phys. Rev.} {\bf D53}
  (1996) 6203--6217}, [\href{http://arxiv.org/abs/hep-ph/9511315}{{\tt
  hep-ph/9511315}}].

\bibitem{Petrelli:1997ge}
A.~Petrelli, M.~Cacciari, M.~Greco, F.~Maltoni and M.~L. Mangano, \emph{{NLO
  production and decay of quarkonium}},
  \href{http://dx.doi.org/10.1016/S0550-3213(97)00801-8}{\emph{Nucl. Phys.}
  {\bf B514} (1998) 245--309}, [\href{http://arxiv.org/abs/hep-ph/9707223}{{\tt
  hep-ph/9707223}}].

\bibitem{Braaten:1994kd}
E.~Braaten and T.~C. Yuan, \emph{{Gluon fragmentation into P wave heavy
  quarkonium}}, \href{http://dx.doi.org/10.1103/PhysRevD.50.3176}{\emph{Phys.
  Rev.} {\bf D50} (1994) 3176--3180},
  [\href{http://arxiv.org/abs/hep-ph/9403401}{{\tt hep-ph/9403401}}].

\bibitem{Ma:1995vi}
J.~P. Ma, \emph{{Quark fragmentation into p wave triplet quarkonium}},
  \href{http://dx.doi.org/10.1103/PhysRevD.53.1185}{\emph{Phys. Rev.} {\bf D53}
  (1996) 1185--1190}, [\href{http://arxiv.org/abs/hep-ph/9504263}{{\tt
  hep-ph/9504263}}].

\bibitem{Braaten:1996rp}
E.~Braaten and Y.-Q. Chen, \emph{{Dimensional regularization in quarkonium
  calculations}}, \href{http://dx.doi.org/10.1103/PhysRevD.55.2693}{\emph{Phys.
  Rev.} {\bf D55} (1997) 2693--2707},
  [\href{http://arxiv.org/abs/hep-ph/9610401}{{\tt hep-ph/9610401}}].

\bibitem{Braaten:1994vv}
E.~Braaten and S.~Fleming, \emph{{Color octet fragmentation and the psi-prime
  surplus at the Tevatron}},
  \href{http://dx.doi.org/10.1103/PhysRevLett.74.3327}{\emph{Phys. Rev. Lett.}
  {\bf 74} (1995) 3327--3330}, [\href{http://arxiv.org/abs/hep-ph/9411365}{{\tt
  hep-ph/9411365}}].

\bibitem{Bodwin:2015iua}
G.~T. Bodwin, K.-T. Chao, H.~S. Chung, U.-R. Kim, J.~Lee and Y.-Q. Ma,
  \emph{{Fragmentation contributions to hadroproduction of prompt$J/\psi$,
  $\chi_{cJ}$, and $\psi(2S)$ states}},
  \href{http://dx.doi.org/10.1103/PhysRevD.93.034041}{\emph{Phys. Rev.} {\bf
  D93} (2016) 034041}, [\href{http://arxiv.org/abs/1509.07904}{{\tt
  1509.07904}}].

\bibitem{Butenschoen:2011yh}
M.~Butenschoen and B.~A. Kniehl, \emph{{World data of J/psi production
  consolidate NRQCD factorization at NLO}},
  \href{http://dx.doi.org/10.1103/PhysRevD.84.051501}{\emph{Phys. Rev.} {\bf
  D84} (2011) 051501}, [\href{http://arxiv.org/abs/1105.0820}{{\tt
  1105.0820}}].

\bibitem{Butenschoen:2012qr}
M.~Butenschoen and B.~A. Kniehl, \emph{{Next-to-leading-order tests of NRQCD
  factorization with $J/\psi$ yield and polarization}},
  \href{http://dx.doi.org/10.1142/S0217732313500272}{\emph{Mod. Phys. Lett.}
  {\bf A28} (2013) 1350027}, [\href{http://arxiv.org/abs/1212.2037}{{\tt
  1212.2037}}].

\bibitem{Chao:2012iv}
K.-T. Chao, Y.-Q. Ma, H.-S. Shao, K.~Wang and Y.-J. Zhang, \emph{{$J/\psi$
  Polarization at Hadron Colliders in Nonrelativistic QCD}},
  \href{http://dx.doi.org/10.1103/PhysRevLett.108.242004}{\emph{Phys. Rev.
  Lett.} {\bf 108} (2012) 242004}, [\href{http://arxiv.org/abs/1201.2675}{{\tt
  1201.2675}}].

\bibitem{Bodwin:2014gia}
G.~T. Bodwin, H.~S. Chung, U.-R. Kim and J.~Lee, \emph{{Fragmentation
  contributions to $J/\psi$ production at the Tevatron and the LHC}},
  \href{http://dx.doi.org/10.1103/PhysRevLett.113.022001}{\emph{Phys. Rev.
  Lett.} {\bf 113} (2014) 022001}, [\href{http://arxiv.org/abs/1403.3612}{{\tt
  1403.3612}}].

\bibitem{Kang:2014tta}
Z.-B. Kang, Y.-Q. Ma, J.-W. Qiu and G.~Sterman, \emph{{Heavy Quarkonium
  Production at Collider Energies: Factorization and Evolution}},
  \href{http://dx.doi.org/10.1103/PhysRevD.90.034006}{\emph{Phys. Rev.} {\bf
  D90} (2014) 034006}, [\href{http://arxiv.org/abs/1401.0923}{{\tt
  1401.0923}}].

\bibitem{Nayak:2005rw}
G.~C. Nayak, J.-W. Qiu and G.~F. Sterman, \emph{{Fragmentation, factorization
  and infrared poles in heavy quarkonium production}},
  \href{http://dx.doi.org/10.1016/j.physletb.2005.03.031}{\emph{Phys. Lett.}
  {\bf B613} (2005) 45--51}, [\href{http://arxiv.org/abs/hep-ph/0501235}{{\tt
  hep-ph/0501235}}].

\bibitem{Nayak:2005rt}
G.~C. Nayak, J.-W. Qiu and G.~F. Sterman, \emph{{Fragmentation, NRQCD and NNLO
  factorization analysis in heavy quarkonium production}},
  \href{http://dx.doi.org/10.1103/PhysRevD.72.114012}{\emph{Phys. Rev.} {\bf
  D72} (2005) 114012}, [\href{http://arxiv.org/abs/hep-ph/0509021}{{\tt
  hep-ph/0509021}}].

\bibitem{Kang:2011mg}
Z.-B. Kang, J.-W. Qiu and G.~Sterman, \emph{{Heavy quarkonium production and
  polarization}},
  \href{http://dx.doi.org/10.1103/PhysRevLett.108.102002}{\emph{Phys. Rev.
  Lett.} {\bf 108} (2012) 102002}, [\href{http://arxiv.org/abs/1109.1520}{{\tt
  1109.1520}}].

\bibitem{Kang:2011zza}
Z.-B. Kang, J.-W. Qiu and G.~Sterman, \emph{{Factorization and quarkonium
  production}},
  \href{http://dx.doi.org/10.1016/j.nuclphysbps.2011.03.054}{\emph{Nucl. Phys.
  Proc. Suppl.} {\bf 214} (2011) 39--43}.

\bibitem{Fleming:2012wy}
S.~Fleming, A.~K. Leibovich, T.~Mehen and I.~Z. Rothstein, \emph{{The
  Systematics of Quarkonium Production at the LHC and Double Parton
  Fragmentation}},
  \href{http://dx.doi.org/10.1103/PhysRevD.86.094012}{\emph{Phys. Rev.} {\bf
  D86} (2012) 094012}, [\href{http://arxiv.org/abs/1207.2578}{{\tt
  1207.2578}}].

\bibitem{Gribov:1972ri}
V.~N. Gribov and L.~N. Lipatov, \emph{{Deep inelastic e p scattering in
  perturbation theory}}, {\emph{Sov. J. Nucl. Phys.} {\bf 15} (1972) 438--450}.

\bibitem{Altarelli:1977zs}
G.~Altarelli and G.~Parisi, \emph{{Asymptotic Freedom in Parton Language}},
  \href{http://dx.doi.org/10.1016/0550-3213(77)90384-4}{\emph{Nucl. Phys.} {\bf
  B126} (1977) 298--318}.

\bibitem{Dokshitzer:1977sg}
Y.~L. Dokshitzer, \emph{{Calculation of the Structure Functions for Deep
  Inelastic Scattering and e+ e- Annihilation by Perturbation Theory in Quantum
  Chromodynamics.}}, {\emph{Sov. Phys. JETP} {\bf 46} (1977) 641--653}.

\bibitem{Baier:2001yt}
R.~Baier, Y.~L. Dokshitzer, A.~H. Mueller and D.~Schiff, \emph{{Quenching of
  hadron spectra in media}},
  \href{http://dx.doi.org/10.1088/1126-6708/2001/09/033}{\emph{JHEP} {\bf 09}
  (2001) 033}, [\href{http://arxiv.org/abs/hep-ph/0106347}{{\tt
  hep-ph/0106347}}].

\bibitem{Gyulassy:2001nm}
M.~Gyulassy, P.~Levai and I.~Vitev, \emph{{Jet tomography of Au+Au reactions
  including multigluon fluctuations}},
  \href{http://dx.doi.org/10.1016/S0370-2693(02)01990-1}{\emph{Phys. Lett.}
  {\bf B538} (2002) 282--288},
  [\href{http://arxiv.org/abs/nucl-th/0112071}{{\tt nucl-th/0112071}}].

\bibitem{Zakharov:1997uu}
B.~G. Zakharov, \emph{{Radiative energy loss of high-energy quarks in finite
  size nuclear matter and quark - gluon plasma}},
  \href{http://dx.doi.org/10.1134/1.567389}{\emph{JETP Lett.} {\bf 65} (1997)
  615--620}, [\href{http://arxiv.org/abs/hep-ph/9704255}{{\tt
  hep-ph/9704255}}].

\bibitem{Baier:1996kr}
R.~Baier, Y.~L. Dokshitzer, A.~H. Mueller, S.~Peigne and D.~Schiff,
  \emph{{Radiative energy loss of high-energy quarks and gluons in a finite
  volume quark - gluon plasma}},
  \href{http://dx.doi.org/10.1016/S0550-3213(96)00553-6}{\emph{Nucl. Phys.}
  {\bf B483} (1997) 291--320}, [\href{http://arxiv.org/abs/hep-ph/9607355}{{\tt
  hep-ph/9607355}}].

\bibitem{Gyulassy:2000er}
M.~Gyulassy, P.~Levai and I.~Vitev, \emph{{Reaction operator approach to
  nonAbelian energy loss}},
  \href{http://dx.doi.org/10.1016/S0550-3213(00)00652-0}{\emph{Nucl.Phys.} {\bf
  B594} (2001) 371--419}, [\href{http://arxiv.org/abs/nucl-th/0006010}{{\tt
  nucl-th/0006010}}].

\bibitem{Wang:2001ifa}
X.-N. Wang and X.-f. Guo, \emph{{Multiple parton scattering in nuclei: Parton
  energy loss}},
  \href{http://dx.doi.org/10.1016/S0375-9474(01)01130-7}{\emph{Nucl. Phys.}
  {\bf A696} (2001) 788--832}, [\href{http://arxiv.org/abs/hep-ph/0102230}{{\tt
  hep-ph/0102230}}].

\bibitem{Wang:200}
P.~B. Arnold, G.~D. Moore and L.~G. Yaffe, \emph{{Photon and gluon emission in
  relativistic plasmas}},
  \href{http://dx.doi.org/10.1088/1126-6708/2002/06/030}{\emph{JHEP} {\bf 06}
  (2002) 030}, [\href{http://arxiv.org/abs/hep-ph/0204343}{{\tt
  hep-ph/0204343}}].

\bibitem{Djordjevic:2003zk}
M.~Djordjevic and M.~Gyulassy, \emph{{Heavy quark radiative energy loss in QCD
  matter}},
  \href{http://dx.doi.org/10.1016/j.nuclphysa.2003.12.020}{\emph{Nucl. Phys.}
  {\bf A733} (2004) 265--298},
  [\href{http://arxiv.org/abs/nucl-th/0310076}{{\tt nucl-th/0310076}}].

\bibitem{Ovanesyan:2011kn}
G.~Ovanesyan and I.~Vitev, \emph{{Medium-induced parton splitting kernels from
  Soft Collinear Effective Theory with Glauber gluons}},
  \href{http://dx.doi.org/10.1016/j.physletb.2011.11.040}{\emph{Phys. Lett.}
  {\bf B706} (2012) 371--378}, [\href{http://arxiv.org/abs/1109.5619}{{\tt
  1109.5619}}].

\bibitem{Sievert:2019cwq}
M.~D. Sievert, I.~Vitev and B.~Yoon, \emph{{A complete set of in-medium
  splitting functions to any order in opacity}},
  \href{http://arxiv.org/abs/1903.06170}{{\tt 1903.06170}}.

\bibitem{Shen:2014vra}
C.~Shen, Z.~Qiu, H.~Song, J.~Bernhard, S.~Bass and U.~Heinz, \emph{{The
  iEBE-VISHNU code package for relativistic heavy-ion collisions}},
  \href{http://dx.doi.org/10.1016/j.cpc.2015.08.039}{\emph{Comput. Phys.
  Commun.} {\bf 199} (2016) 61--85},
  [\href{http://arxiv.org/abs/1409.8164}{{\tt 1409.8164}}].

\bibitem{Isgur:1989vq}
N.~Isgur and M.~B. Wise, \emph{{Weak Decays of Heavy Mesons in the Static Quark
  Approximation}},
  \href{http://dx.doi.org/10.1016/0370-2693(89)90566-2}{\emph{Phys. Lett.} {\bf
  B232} (1989) 113--117}.

\bibitem{Isgur:1989ed}
N.~Isgur and M.~B. Wise, \emph{{WEAK TRANSITION FORM-FACTORS BETWEEN HEAVY
  MESONS}}, \href{http://dx.doi.org/10.1016/0370-2693(90)91219-2}{\emph{Phys.
  Lett.} {\bf B237} (1990) 527--530}.

\end{thebibliography}\endgroup

\end{document}